\documentclass[useAMS,usenatbib]{mn2e}
\usepackage{graphicx,times}
\usepackage{amssymb}

\newcommand{\galfit}{\textsc{Galfit}}

\newcommand{\sex}{\textsc{SExtractor}}
\newcommand{\gdd}{\textsc{Gim2d}}
\newcommand{\dt}{\textsc{DrizzlyTim}}
\newcommand{\aw}{\textsc{Astro-wise}}
\def\Reff{\ifmmode{R_\mathrm{e}}\else{$R_\mathrm{e}$}\fi}
\def\mueff{\ifmmode{\langle \mu \rangle _{\rm e}}\else{$\langle \mu \rangle _{\rm e}$}\fi}

\title[Structural parameters of galaxies in the Coma cluster line of sight.]{The HST/ACS Coma Cluster Survey III.\\ 
Structural Parameters of Galaxies using single-S\'ersic Fits.
\thanks{Based on observations made with the NASA/ESA Hubble
Space Telescope, obtained at the Space Telescope Science Institute,
which is operated by the Association of Universities for Research in
Astronomy, Inc. under  NASA contract  NAS 5-26555. These observations
are associated with programme GO10861.}}

\author[Carlos Hoyos et al.]{Carlos Hoyos$^{1,2,3}$, Mark den Brok$^{4}$, Gijs Verdoes Kleijn$^{4}$, David Carter$^{7}$,\newauthor
 Marc Balcells$^{5,6}$, Rafael Guzm\'an$^{1}$, Reynier Peletier$^{4}$, Henry C. Ferguson $^{9}$,\newauthor 
 Paul Goudfrooij$^{9}$, Alister W. Graham$^{11}$, Derek Hammer$^{8}$, Arna M. Karick$^{7}$,\newauthor 
 John R. Lucey$^{10}$,  Ana Matkovi\'c$^{9,12}$, David Merritt$^{13}$, Mustapha Mouhcine$^{7}$\newauthor
 Edwin Valentijn$^{4}$
\\$^{1}$Department of Astronomy, University of Florida, PO Box 
112055, Gainesville, FL 32611, USA.
\\$^{2}$Departamento de F\'isica Te\'orica, Facultad de Ciencias, Universidad Aut\'onoma de 
Madrid, Cantoblanco, 28049 Madrid, Spain
\\$^{3}$School of Physics and Astronomy, The University of Nottingham,
University Park, Nottingham, NG7 2RD, UK.
\\$^{4}$Kapteyn Astronomical Institute, University of Groningen, 
PO Box 800, 9700 AV Groningen, The Netherlands.
\\$^{5}$Instituto de Astrof\'isica de Canarias, C/V\'ia Lactea s/n, 
38200 La Laguna, Tenerife, Spain.
\\$^{6}$Isaac Newton Group of Telescopes, Apartado de Correos 321, 
E-38700 Santa Cruz de la Palma, Canary Islands, Spain. 
\\$^{7}$Astrophysics Research Institute, Liverpool John Moores 
University, Twelve Quays House, Egerton Wharf, Birkenhead CH41 1LD, UK.
\\$^{8}$Department of Physics and Astronomy, Johns Hopkins 
University, 3400 North Charles Street, Baltimore, MD 21218, USA.
\\$^{9}$Space Telescope Science Institute, 3700 San Martin Drive, Baltimore, MD 21218, USA.
\\$^{10}$Department of Physics, Durham University, South Road, Durham,
DH1 3LE, UK.
\\$^{11}$Centre for Astrophysics and Supercomputing, Swinburne
University of Technology, PO Box 218, Hawthorn, VIC 3122, Australia.
\\$^{12}$Astronomy and Astrophysics, Pennsylvania State University,
525 Davey Lab, University Park, PA 16802, USA.
\\$^{13}$Department of Physics and Center for Computational Relativity and Gravitation, 
Rochester Institute of Technology, Rochester, NY 14623, USA.
}

\begin{document}

\date{}

\pagerange{\pageref{firstpage}--\pageref{lastpage}} \pubyear{2009}

\maketitle

\label{firstpage}

\begin{abstract}
We present a catalogue of structural parameters for 8814 galaxies in the 25 fields of the HST/ACS
Coma Treasury Survey.  Parameters from S\'ersic fits to the two-dimensional surface brightness 
distributions are given for all galaxies from our published Coma photometric catalogue with
mean effective surface brightness brighter than 
$26.0~\mathrm{mag}~\mathrm{arcsec}^{-2}$ and brighter than
24.5 mag (equivalent to absolute magnitude - 10.5), 
as given by the fits, 
all in F814W(AB).  
The sample comprises a mixture of Coma members and background objects; 
424 galaxies have redshifts and of these 163 are confirmed members. 
The fits were carried out using both the \gdd\ and \galfit\ codes. 
We provide the following parameters: Galaxy ID, RA, DEC, the total corrected automatic magnitude
from the photometric catalogue, 
the total magnitude of the model ($\mathrm{F814W}_{\mathrm{AB}}$), 
the geometric mean effective radius {\Reff}, the mean
surface brightness within the effective radius $\langle \mu \rangle _{\rm e}$, the S\'ersic index $n$, the 
ellipticity and the source position angle.   
The selection limits of the catalogue and the errors listed for the S\'ersic parameters come 
from extensive simulations of the fitting process using synthetic galaxy models.  
The agreement between \gdd\ and \galfit\ parameters is sensitive to details of the fitting procedure; for the 
settings employed here the agreement is excellent over the range of parameters covered in the catalogue.  
We define and present two goodness-of-fit indices which quantify the degree to which the image can be
approximated by a S\'ersic model with concentric, 
coaxial elliptical isophotes; such indices may be used to objectively select galaxies with more complex 
structures such as bulge-disk, bars or nuclear components. 

We make the catalog available in electronic format at Astro-WISE and MAST.
\end{abstract}


\section{Introduction}

Surface brightness distributions are a vital tool in our understanding of galaxies. Since the pioneering work of 
Reynolds (1913) and Hubble (1930) on elliptical galaxies it has become common to
fit the radial surface brightness distributions to functions having a small number of parameters, which include
a scale length, a characteristic surface brightness, and one or two further parameters which describe the 
structure of the surface brightness profile. The most commonly used fitting function is that of 
S\'ersic (1963, 1968), whose function includes as special cases both the
$R^{1/4}$ law of de Vaucouleurs (1948) and the exponential surface
brightness distribution which is characteristic of disk galaxies (Patterson 1940; de Vaucouleurs 1957, 1959; Freeman 1970):
\begin{equation}
I(R) = I_e\,\exp\{-b\,\left[(R/\Reff)^{1/n} - 1\right]\},
\end{equation}
\noindent where $I(R)$ is the specific intensity at distance $R$ from the centre, \Reff\ is the radius 
enclosing half the galaxy light, $I_e$ is the specific intensity at \Reff, $n$ is the S\'ersic index or 
concentration index (Trujillo et al.~2001), and $b \approx 1.9992\times n -0.3271$ (Capaccioli 1989).  

The S\'ersic function provides a good model for ellipticals, giants showing values of $n \geq 4$, 
intermediate luminosity ellipticals $n \approx 2-4$
and dwarfs $n \approx 1-2$ (Caon, Capaccioli \& D'Onofrio 1993; Graham et al. 1996; Graham \& Guzm\'an 2003).  
Bulges of disk galaxies are also well fit by the S\'ersic model (Andredakis, Peletier \& Balcells 1995) 
with indices $n \approx 0.5 - 4$ (Balcells, Graham \& Peletier 2007; Graham \& Worley 2008). For disk galaxies, pure S\'ersic 
fits often yield poor approximations to the entire galaxy surface brightness distribution, due to the 
presence of bulges, bars, spirals, outer disk truncations  (e.g. van der Kruit \& Searle 1981a,b) and 
anti-truncations (Erwin et al.~2005). However, classifying galaxies into early ($n > 2.5$) and late 
($n< 2.5$) types on the basis of single S\'ersic fits to the entire galaxy has become standard practice 
(e.g. van der Wel et al.~2008), especially in samples with limited image depth and spatial resolution 
which prevent more complex modeling.  This practice fails when the samples include lower luminosity dwarf
elliptical galaxies. The reliability of such fits may be calibrated by performing 
single S\'ersic fits to nearby, well resolved galaxies. Needed for the interpretation of such single 
S\'ersic fits is a parameter that quantifies the degree to which the true surface brightness distribution 
deviates from the S\'ersic model.  

The HST/ACS Treasury Survey of the Coma cluster was presented in Carter et al.\ (2008, Paper I).  
Although the survey was originally planned to cover 740 arcmin$^2$ of the Coma cluster field, the final 
areal coverage  is 274 arcmin$^2$ in the F475W and F814W bands, mostly in the core region, owing to the ACS failure in 2007 January.
Still, with the exquisite quality and depth of the imaging and the large number of spectroscopic redshifts 
known for galaxies in this field
(Colless \& Dunn 1996; Mobasher et al.~2001; Marzke et al.~2010; Chiboucas et al.~2010) this survey 
allows studies of the structure of large samples of cluster members to an unprecedented depth.  
The photometric catalogue from the HST/ACS images was presented in 
Hammer et al.\ (2010, Paper II; see Sect.~\ref{sec:InputData}).  

This paper presents a structural analysis of sources selected from a structural analysis of the sources from the Paper~II photometric catalogue,  
based on two-dimensional single S\'ersic fits.  Of the $\sim$75,000 objects in that catalog, we provide 
S\'ersic parameters for 8814 galaxies that are located both inside the cluster and in the background; the selection function 
is explained in Sect.~\ref{cut:mag}. We present standard S\'ersic parameters as well as two goodness-of-fit 
indices, that provide a quantitative measure of the degree to which the galaxy surface brightness distribution 
deviates from a S\'ersic model with concentric, co-axial elliptical isophotes (Sect.~\ref{sec:goodnessoffit}).
These indices can be used to identify those galaxy images which allow for additional components, such as 
outer disks, nuclear components or bars.  Given the complexity of the structural analysis process, we focus 
this paper on the presentation of the analysis techniques and of the catalogue.  We defer scientific analysis 
to future papers.  The structural parameters presented here can be used:

\begin{itemize}
\item{To study the cosmological evolution of galaxy sizes and shapes by using the Coma cluster as a local reference sample.}
\item{To quantify the faint end of global scaling relations, such as size-(surface brightness) diagrams and the Fundamental Plane,
revealing how dwarf elliptical galaxies do or do not unite with brighter ellipticals.}
\item{To study the correlation between the structural parameters and the photometric masses of elliptical and 
lenticular galaxies, which could be used in cluster membership studies (Trentham et al.~2010).}
\end{itemize}

Our results in Coma can be compared with the lower density Virgo and Fornax cluster environments
where targeted HST/ACS surveys provide structural information at higher physical resolution for smaller 
samples of galaxies (Ferrarese et al. 2006; C\^ot\'e et al. 2007).
The Coma data set may be also used in conjunction with HST surveys at higher redshift to study 
the evolution of the structural properties of galaxies. STAGES (Gray et al.~2009) is a
survey of the supercluster Abell 901/2 at a redshift of 0.165. Amongst an extensive multi-wavelength
dataset, ACS images have been used for S\'ersic fits to a large sample of galaxies in the STAGES region. 
GEMS (Rix et al.~2004) is an ACS survey of a 900 arcmin$^2$ region within the 
Extended Chandra Deep Field South region. Although it is a field rather than cluster survey, 
it provides a useful evolution benchmark at redshifts approaching $z=1$. HST has been used to 
study the structural properties of galaxies in higher redshift clusters, where there is a suggestion
of size evolution by up to a factor four (e.g. Trujillo et al. 2006; Strazzullo et al. 2010). 

There are currently a number of codes capable of performing two-dimensional S\'ersic model fits
to the surface brightness distribution of galaxies. Two extensively used  
are \gdd\ (Simard et al.~2002) and \galfit\ (Peng et al.~2002).
Both codes work by minimising a merit function, and produce similar outputs, 
but their inner workings differ in a number of ways, such as the minimisation technique: 
\gdd\ uses the Metropolis algorithm, whereas \galfit\ uses the Levenberg-Marquardt algorithm. 
\galfit\ offers practical advantages, such as higher execution speed and the ability 
to simultaneously fit several targets.  But because each code has its own merits, we carried out the fits using both codes, and present both results.  In order not to bias the comparison, two teams worked largely independently, one with \gdd, and another with \galfit. 
While some details differ, e.g. in the parameter ranges explored in the Montecarlo simulations, there was enough coordination to ensure the results would be comparable to each other.  
We show in Sect.~\ref{sec:comparison} that the agreement is very good.   

\gdd\ and \galfit\  were compared by
H\"aussler et al.~(2007), who concluded that both could produce similar 
results, but warned against a systematic underestimate of both the total luminosity and effective 
radius of the \gdd\ output for lower surface brightness sources. 
We were able to reproduce their findings; in Sect.~\ref{sec:setup_G2D} we present an approach which 
successfully overcomes these biases. 

The paper is structured as follows. \S~\ref{sec:InputData} describes the input data from which our
catalogue is derived. \S~\ref{sec:setup_G2D} and \S~\ref{sec:setup_galfit} describe how the 
\gdd\ and \galfit\ analysis runs on the Coma data were setup. \S~\ref{sec:goodnessoffit} introduces two 
additional parameters calculated from the residuals of the data from the models which describe 
how well the S\'ersic model fits the data. \S~\ref{sec:catalogue} presents the final structural parameters
catalogue and the criteria for inclusion in the catalogue. 
\S~\ref{sec:comparison} presents a comparison between the results obtained with two codes on 
the Coma data. 
In \S~\ref{sec:conclusions} we present our
conclusions and describe the next steps in the analysis
of the galaxies and their surface brightness distributions. In
Appendix~\ref{apx:g2d_gfit07} we compare our results with those of
H\"aussler et al (2007). 

Throughout the paper we assume the distance to Coma of 100 Mpc, corresponding to a distance 
modulus of $m-M = 35.0$ (see Paper~I).  All magnitudes are in the AB system.  In this paper, with a few exceptions, we express
the surface brightness in terms of the mean effective surface brightness {\mueff}, i.e., the mean surface brightness enclosed within {\Reff}. Graham
\& Driver (2005) show that the relationship
between \mueff\ and the effective surface brightness $\mu_{\rm e}$ (surface brightness \textsl{at} \Reff) is given by: 
\begin{equation}
\langle\mu\rangle_{\rm e} = \mu_{\rm e} -2.5\log [f(n)],
\end{equation}
\noindent where:
\begin{equation}
f(n)=\frac{n {\rm e}^{b}}{b^{2n}}\int_{0}^{\infty}{\rm e}^{-x}x^{2n-1} {\rm d}x=\frac{n {\rm e}^{b}}{b^{2n}}\Gamma (2n),
\end{equation}
\noindent and $\Gamma (2n)$ is the complete gamma function. Total magnitude $m$ is related to \mueff\
by the simple relation:
\begin{equation}
m = \langle\mu\rangle_{\rm e} - 2.5\log(2\pi R_{\rm e}^2),
\end{equation}

\section{Data}
\label{sec:InputData}

\subsection{HST/ACS images}
\label{sec:Images}

The survey design and reduction of the images are described in detail in Paper I, so only a summary 
will be provided here.  The observations (program GO 10861) were taken between 2006 November and 
2007 January with the HST/ACS camera (2$\times$4096$\times$2048 pixels, 0.05 arcsec/pixel, 3 arcsec 
interchip gap).  A total of 25 visits were completed before the failure of ACS.  Most fields (19/25) are 
located within 0.5 Mpc of the cluster centre (the full list of survey fields is given in Table~2 of Paper~I). 
The remaining six fields are in the South-West extension of Coma. Two HST orbits were devoted to each 
pointing.  A four-position dither pattern was used for each of the F475W and F814W images, with total 
integration times of 2560 s and 1400 s, respectively. The dither pattern allowed us to fill the ACS inter-chip 
gap, albeit with lower S/N. Total exposure times were lower for some visits due to dither positions that failed to 
acquire guide stars. Final exposure times are given in Table~5 of Paper~I.  

Data reduction was carried out with a dedicated pipeline. It included the combination of individual images 
with the Multi-Drizzle software (Koekemoer et al.~2003), which yields combined images resampled onto a 
rectified (but original sky orientation) output frame with 0.0495 arcsec/pixel. Cosmic rays were removed 
during the multi-drizzle process and also using \textsc{lacosmic} (van Dokkum 2001).  These processed 
images, together with an initial source catalogue comprised the first data release (DR1), 2008 
August\footnote{MAST (archive.stsci.edu/prepds/coma/) and Astro-WISE (www.astro-wise.org/projects/COMALS/)}.  

\subsection{Catalogues of the Coma Data Release 2}
\label{sec:catalogs}

The second data release (DR2), available at the same web sites as DR1, includes improvements in 
alignment between F814W and F475W images, better astrometry, aperture corrections to the \sex\ 
photometry, and photometry of sources that project onto large galaxies.  Details of the data processing 
and description of the DR2 photometric catalogues, including the \sex\ (version 2.5; Bertin \& Arnouts 1996) 
configuration parameters employed in the catalogue generation, are given in Paper~II.  The DR2 catalogues 
contain $\sim$73,000 sources. Based on Monte-Carlo simulations, the 80\% completeness limit for point sources 
in the DR2 catalogues is 27.8 mag in F475W and 26.8 mag in F814W. 

The DR2 images and catalogues are the basis for the structural analysis done with \galfit, whereas \gdd\ fitting 
was performed on the DR1 images and catalogue. This difference represents no problems. 
Comparison of the \sex\ catalogues from the two releases shows
that 99\% of the detected sources match, with 75\% of the additional
catalogue objects in DR1 being in Visit~03, which lacked two of
the four dither positions. These additional objects would in any case
be fainter and smaller than the catalogue limits of the current paper (Sect.~\ref{cut:mag}).  
We refer to the DR2 catalogue as the ``photometric catalogue'' throughout this paper.  

\subsection{PSF}
\label{subsec:psf}

The Point Spread Function (PSF) is a key ingredient of any signal-to-noise weighted analysis 
of the morphological and structural properties of galaxies. 
The PSF of the Wide Field Channel (WFC) of the ACS has been extensively
studied. Jee et al.~(2007) and Rhodes et al.~(2007) offer different
suites for creating ACS PSFs for a variety of observing conditions.
The HST Intrument Science Report 0306 (Krist 2003) presents a detailed study of the variation of the PSF across 
the WFC chips. The PSF
of the WFC depends both on time and position on 
the chip. The \textsc{TinyTim} program (Krist 1993) takes advantage of this
empirical knowledge, and creates artificial PSFs for a large variety of
observing conditions and HST instruments.

We created a grid of ACS PSFs using \textsc{TinyTim}. These were then
combined using the code \textsc{DrizzlyTim}, by Luc Simard, kindly made available 
to us by the author.  

\textsc{DrizzlyTim} calculates the location
of the original PSFs in the calibrated flat fielded individual
exposures, using the same multidrizzling parameters and shift file as used to produce the science images.
\textsc{DrizzlyTim} then invokes \textsc{TinyTim} to create the required PSFs in the calibrated, flat fielded set of coordinates.
The PSFs are created with an oversampling of 5, and assuming a 6500 K black body as a representation of the 
object spectrum. This is appropriate for the E and S0 galaxies in our sample.
\textsc{DrizzlyTim} then places these PSFs into blank frames, with the same size and header parameters
as those of the real flat fielded individual exposures. These frames are then coadded, using again
the same multidrizzle parameters as those used to manufacture the final science images.
The final step is to apply the Charge Diffusion Kernel to these newly created PSFs in the calibrated, geometrical distortion
corrected images. We followed this process to create a grid of PSFs with the sampling of the original images. One PSF was
created every 150 pixels in  $x$ and $y$ directions, and
the PSF imagelets created were 31 pixels on a side. The typical FWHM
of the PSFs created was 2.0 pixels, with at most a 20\% variation 
across the field. As the time period over which the images were obtained was short, we 
did not allow for temporal variation of the PSF.
When fitting real galaxies, both \gdd\ and \galfit\ were instructed to use the nearest PSF 
to that object. 

\begin{figure}
\begin{center}
\includegraphics[scale=0.25]{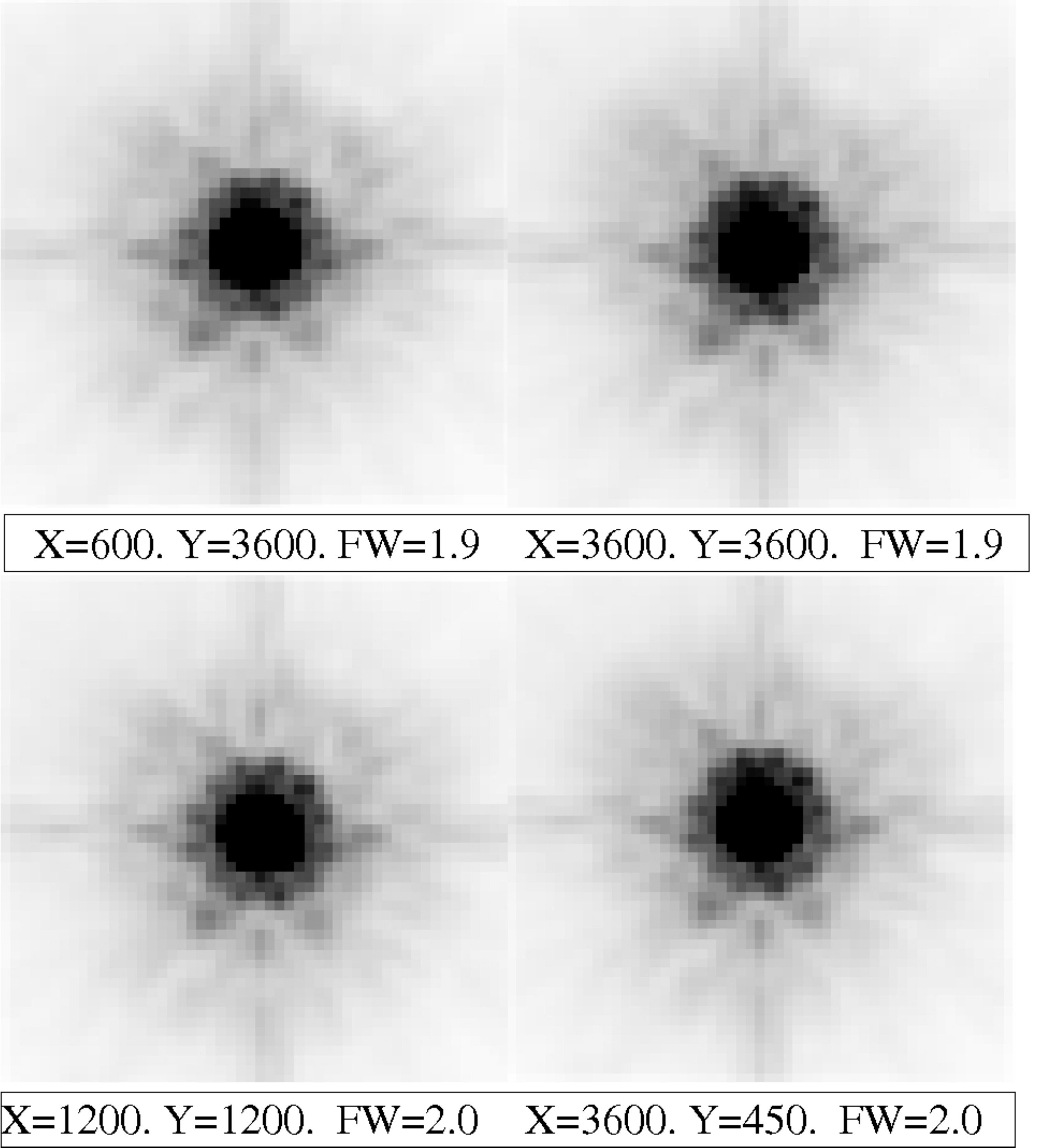}
\caption{Four examples of the PSFs created by DrizzlyTim. These PSFs were used by \gdd\ and \galfit\ to fit the
real sources in the science images. 
The image in each panel is 50 pixels (2.0 arcseconds) square.}
\label{psfshow}
\end{center}
\end{figure}

Fig.~\ref{psfshow} shows a set of \textsc{DrizzlyTim} PSFs, with their respective FWHM.

\section{Gim2d fitting}
\label{sec:setup_G2D}

This section describes the \gdd\ setup used to fit the surface brightness distributions
of the detected galaxies. As will also be done in Sect.~\ref{sec:setup_galfit} for \galfit, we build 
upon the experience of the GEMS collaboration who analyse galaxies from HST/ACS imaging, 
and perform extensive simulations to compare systematics in \gdd\ and \galfit\ (H\"aussler et al.~2007).  
From our own analysis of the H\"aussler et al.~paper, shown in Appendix~\ref{apx:g2d_gfit07}, we 
conclude that mask creation is critical for \gdd, and we present a new prescription for creating the 
masks that \gdd\ requires for object detection and fitting.  
In addition, this section presents MonteCarlo simulations used to assign statistical errors to 
the \gdd\ best-fit parameters. 

\subsection{Object masks for \gdd}
\label{sec:object_masks_gdd}

Our own simulations, and those of H\"aussler et al.~(2007), show that \gdd\ can miss a substantial 
fraction of light from faint sources (a comparison of our simulations to those of H\"aussler et al.~is 
given in Appendix~\ref{apx:g2d_gfit07}).  Our analysis concludes that this problem originates in the 
use of \sex\ segmentation masks as input for \gdd.  When \gdd\ is instructed to infer its initial 
guesses from \sex\ magnitude and size parameters (\texttt{DOINIT}=\texttt{YES}), and is allowed to 
refine the sky level estimate obtained from \sex\ (\texttt{DOBKG}=\texttt{YES}), the use of \sex\ 
segmentation masks leads to systematically fainter and smaller solutions.  Setting \texttt{DOBKG} 
and \texttt{DOINIT} to \texttt{NO}
(i.e. sky background fixed from {\sex}, and intitial estimates of other parameters
taken from the parameter file {\it gal.mdpar}) fixes the systematic error in total magnitude and \Reff, but at the expense of 
an increase scatter in the solutions, and a dramatic increase in the convergence time.  
When \texttt{DOBKG} and \texttt{DOINIT} are both set to \texttt{YES}, \gdd\ is left free to automatically 
decide which section of the global parameter space to explore. 
\gdd\ does this using the mask image it has been provided. It first estimates the sky using 
the pixels designated as sky pixels in the mask image that lie 
a fixed number (which we set to 10) of pixels away from the target mask. It then subtracts 
this estimate of the sky value from the input image, and derives 
initial estimates of the total flux, inclination angle, ellipticity
and effective radius. In this important step, \gdd\ calculates the total flux and the effective radius of the
target using pixels which, according to the mask it has been given,
belong to the \texttt{ISOAREA} of that object.
\gdd\ will then explore the range in parameter space from 0 counts to twice the sky subtracted flux 
within the mask, and from 0 pixels to twice the effective radius of
the set of pixels within the mask. This factor of two is 
hard coded into \gdd\ and can not be tweaked.

Therefore if \gdd\ is
set with both \texttt{DOBKG=YES}, \texttt{DOINIT=YES} and fed with
a segmentation image from \sex, it will only explore
magnitudes between (\texttt{MAG\_ISO} - 0.75) and infinity.
However, the  \sex\ simulations presented in Paper II show 
that for the fainter sources the real magnitudes can be off 
from  \texttt{MAG\_AUTO} by up to two magnitudes. A very similar statement could be made
for the effective radius. In this case, the range in linear size
explored is from 0 pix to twice the \gdd\ 
initial estimate, which is built from the two dimensional Kron
radius. This effective radius estimate was found to be 
different from  the true effective radius by up to a factor of 5 for
the less luminous sources. 

This clearly indicates that the masks \gdd\ is 
provided have to be enlarged, if they are to be a faithful
representation of the real extent of the targets.
Instead of using the standard segmentation image created by {\sex}, we
build a customized mask for each object, using the information 
from the \sex\ catalogue for the whole frame and the knowledge
of the noise properties of the images.
This mask image is constructed separately for each particular object, since \gdd\ treats target and background sources 
differently. Any given object is represented by one aperture when acting as the target, and is represented
by another smaller aperture when being considered a background source possibly affecting the fit of a different object. 
The properties of the proposed masks, together
with the practical steps required to create them, are summarized in Appendix~\ref{apx:mask_recipe}.  

Fig.~\ref{figuramascaras} shows four examples of galaxies, together
with their associated masks. Black pixels belong to the target object,
white pixels belong to other sources, and grey pixels are sky. The 
first object is a low surface brightness source that was however 
detected by {\sex}. The second and fourth objects are spiral
galaxies. In all cases, the FOVs
are given in the image insets. 
The Kron-like apertures which were adopted for the targets, presented as black pixels on the mask images, 
are clearly more extended than the visible flux from the target galaxy, and are typically much larger than 
the isophotal apertures used for neighboring objects which are shown as white pixels.
The apertures for the background sources can overlap the aperture of the
target object, which is not possible when using the segmentation
images produced by {\sex}, when the background objects can potentially
interfere with the ability of \gdd\  to measure regions outside
the \texttt{ISOAREA\_IMAGE} of the target. 

\begin{figure}
\begin{center}
\includegraphics[width=8cm]{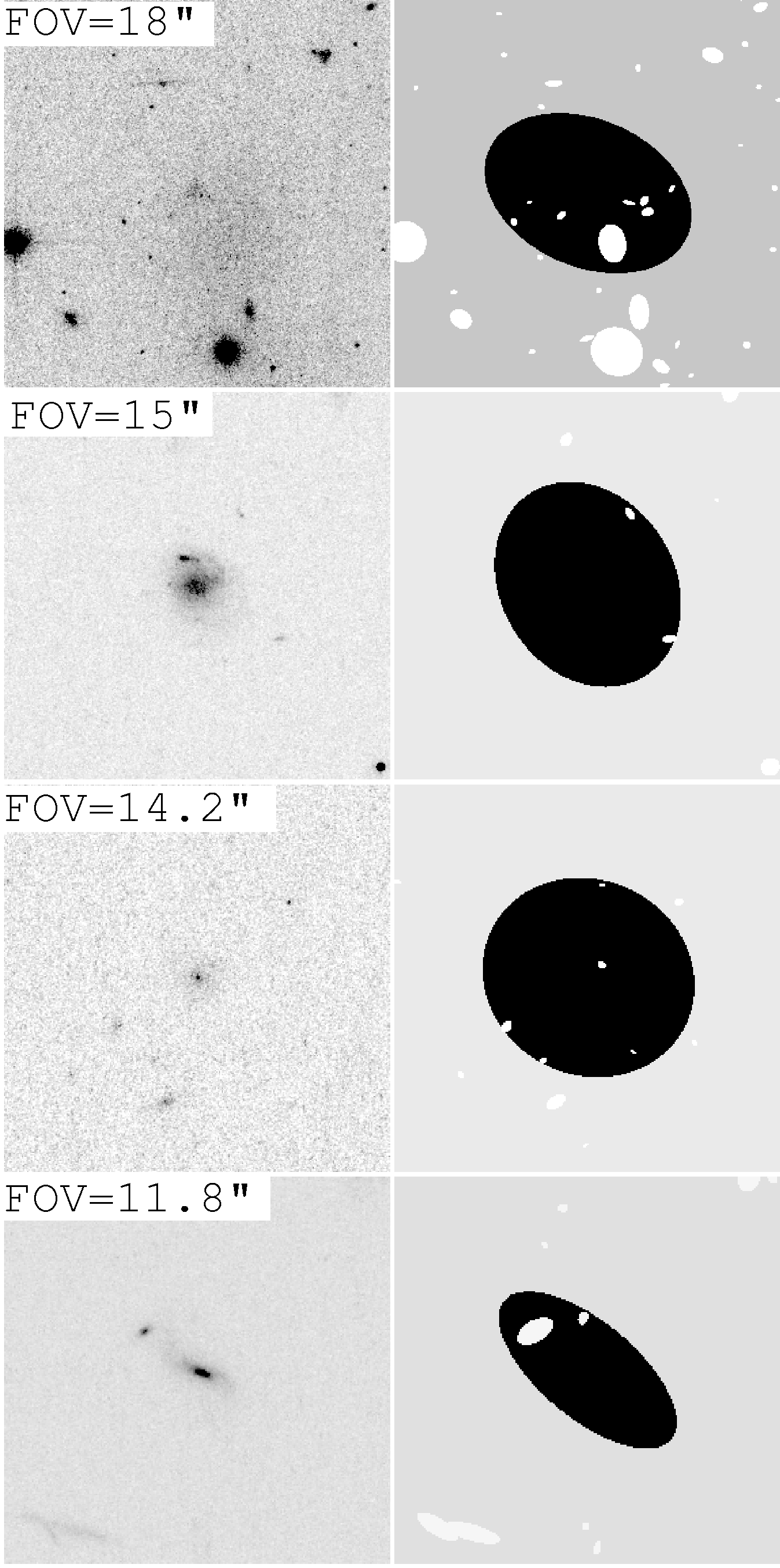}
\caption{Examples of the masks used by {\gdd}. The ``Kron-like'' apertures devised for the 4 
target objects are much larger than the apertures of their
neighboring sources, which have an area equal
to their \texttt{ISOAREA\_IMAGE}.}
\label{figuramascaras}
\end{center}
\end{figure}

\subsection{Noise model}

A simple noise model is used for the fits. When \gdd\ is not given a specific noise image, it 
builds an internal weight map based upon the \textsl{rms} of the 
background ($\sigma_{\mathrm{bkg}}$) and  the effective gain, which 
is a function of the effective exposure time of the science frames. This noise model is a transalation of the
usual CCD uncertainty equation. In our case, $\sigma_{\mathrm{bkg}}$ is taken directly from 
the \sex\ catalogue, although it is later recalculated by \gdd\ in most cases. The effective exposure time is 
read from the header of the HST/ACS frames.

With \texttt{DOBKG} set to \texttt{YES},  \gdd\ refines the sky value
given by {\sex}, and  obtains a better estimate of  $\sigma_{\mathrm{bkg}}$. This is then used to construct
the internal weight map. The sky pixels involved in this calculation were at least 10 pixels away from those pixels 
determined to belong to the target object. The refined sky value is the median value of at least 30 pixels, applying a $5\sigma$ 
clipping thresholding scheme. The \textsl{rms} of the background obtained by \gdd\ in this way is 
generally in excellent agreement with the $\sigma_{\mathrm{bkg}}$ initially estimated by {\sex}, both in the mean and variance.
In cases where insufficient numbers of pixels were available to estimate the background rms inside the imagelet (owing to 
crowded fields), GIM2D defaulted to the user-supplied rms that was estimated by SExtractor.

\subsection{Final {\gdd} configuration}

\texttt{DOINIT} was set to \texttt{NO}. The lower limit of the total
flux was set to 0, while the upper limit was set to 10 times the automatic aperture flux.
The Bulge-to-Total fraction was set to 1. The effective radius ranged 
 from 0.0 to 10.0 times the effective radius estimate obtained for a pure S\'ersic model of $n=2.25$ of the same 
\texttt{MUOBS} and \texttt{FLUX\_RADIUS}.
The ellipticity and position angle were allowed to search their whole ranges.
The \textsl{X} and \textsl{Y} drifts were permitted to range from  $-10$ to $10$ pixels, and the residual sky value 
was allowed to go from $-0.25$ to $0.25$. The S\'ersic index $n$ was allowed
to range from 0.25 to 10.0. Thus, each object had an individualized \gdd\ configuration file leading to all objects
being fit by a single S\'ersic profile.
The Metropolis temperatures are adequate for the explored ranges, and expected typical changes in each iteration. 
In all cases, saturated pixels were rejected from the fits. The Metropolis
algorithm is given 400 iterations to cool off after achieving convergence (see Simard et al. 2002 for more technical detail). 

\subsection{Errors on the parameters and depth of the survey}
\label{g2d:simulations}

Although \gdd\ produces, together with its results, a set of
confidence intervals for the fitted parameters, the error estimates
represent only the scale upon which the Figure-of-Merit that \gdd\
uses is expected to vary. Therefore these confidence intervals merely
reflect how constrained the fit is. A more realistic and meaningful
error analysis needs to investigate the extent to which the minimum of the Figure-of-Merit
can drift in its parameter space. 

A modest number of Monte Carlo \gdd\ simulations was run. The purpose
of these simulations is twofold. The first is 
to be able to ascribe realistic statistical errors to the fits
produced by {\gdd}. The second purpose is to assess the surface 
brightness limit beyond which it will not be possible to recover reliable 
structural parameters.

10,000 model images were created using \textsc{galimage} 
within \textsc{IRAF.Fuzzy}. In 
this step, a Moffat (1969)  PSF representative of the average
properties of several non-saturated stars  
was used to degrade the galaxy models. \textsc{mknoise}  was then used 
to add appropriate Poissonian noise to this model. This 
blurred image was then added to a real ACS image which 
therefore provides the readout noise, and the bulk of the 
error correlation that is typical of ACS drizzled images. 
The selected canvas images correspond to
visits 1, 15, 78, and 90 (see Paper I). \sex\ and \gdd\ 
were run, with the same parameters, weight images 
and flag images as those used to create the \sex\ catalogue and 
the same  experimental setup described above, using the Moffat
PSF as the convolution kernel.

The mean effective surface brightnesses of the
models ran from 19.0 to 27.0, 
with effective radii distributed randomly in $\log{\Reff}$ between 2.0 and 60.0 pixels. 
S\'ersic indices were  randomly 
distributed between 0.5 and 4.5. 
Ellipticities were randomly distributed 
from 0.0 to 0.8 and position angles were unconstrained.
Of the total number of model galaxies created, 10,000 were 
both detected by \sex\ and successfully analyzed by \gdd; 
the analysis presented in this subsection deals with these 10,000 fake sources.
The remaining sources were either not detected by \sex\ or
fell in problematic areas of the image such as the 
CCD edges and were thus rejected.

\begin{figure*}
\begin{center}
\includegraphics[scale=0.30]{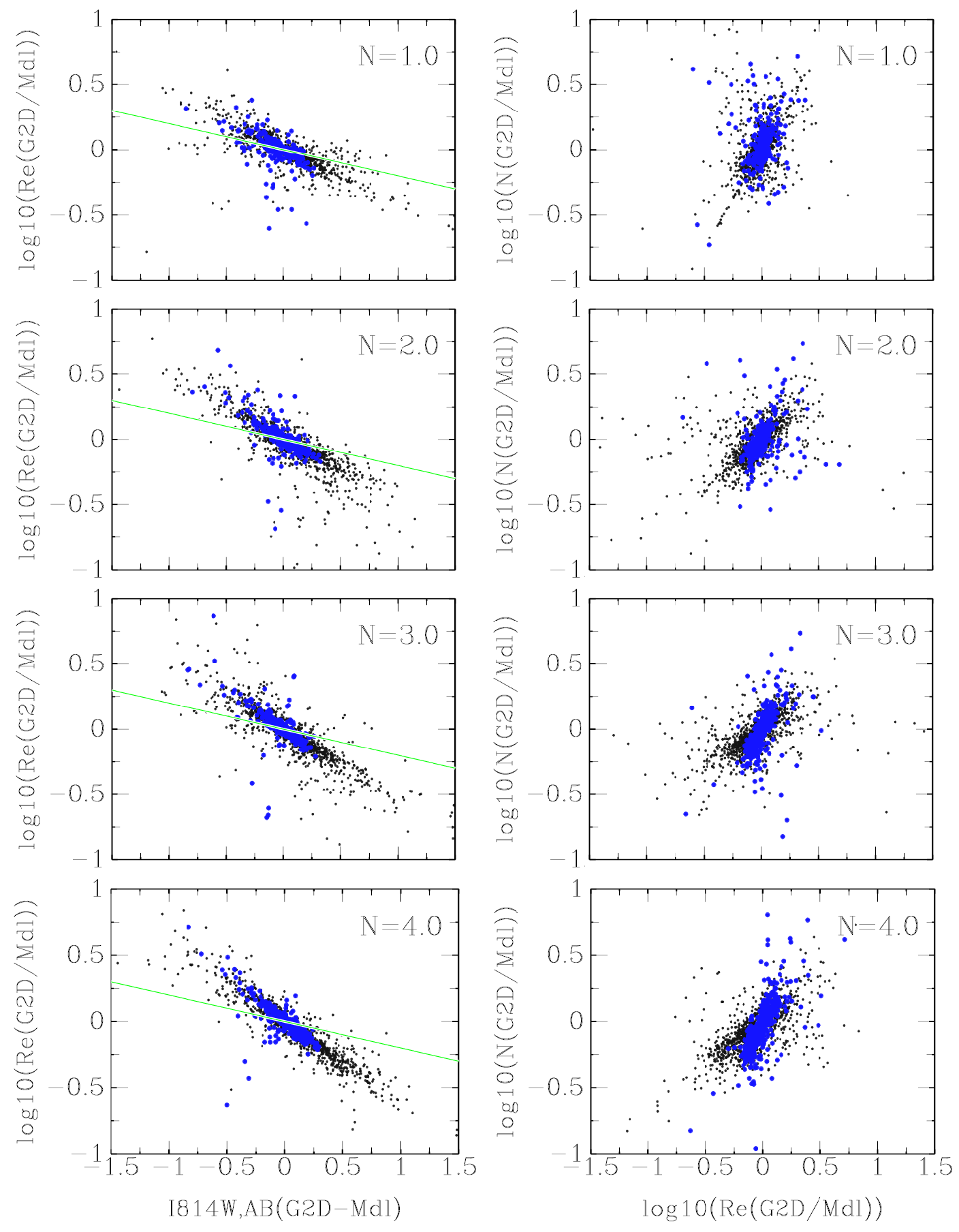}
\caption{Effective radius residuals \textsl{vs.} magnitude residuals
  (left panels) and S\'ersic index residuals \textsl{vs.} effective
  radius residuals (right panels)  as a function of the input S\'ersic index and {\mueff}. 
The residuals are defined as the ratio of the output value from \gdd\ to the input model
value. 
From top to bottom, the panels include sources with input S\'ersic index 
$0.5 < n < 1.5$, $1.5 < n < 2.5$, $2.5 < n < 3.5$, and $3.5 < n < 8.0$, respectively. 
The blue dots represent points with input \mueff\
brighter than 22.0, and the black dots represent the whole pool of
models created. The green line shows the location of the points
should the fitting process preserve the 
mean effective surface brightnesses of the
models. Deviations from this line indicate that \gdd\ was not able to
accurately retrieve the value of \mueff. Also, the slope of the clouds is correlated with
the intrinsic profile of the model being fit. For galaxies with large
$n$, it is somewhat  more difficult to reproduce the parameters of
the input model. The right set of panels shows how the fits to the
models expand or compress, depending on whether
the fit overestimates or underestimates $R_\mathrm{e}$.}
\label{fig:g2d_simulations_1}
\end{center}
\end{figure*}

The first step in the analysis of these simulations is presented in 
Fig.~\ref{fig:g2d_simulations_1}. This figure shows the 
magnitude residuals against the effective radius residuals, and the effective
radius residuals against the S\'ersic index residuals. These relations
are presented in different panels,
according to the input $n$ as indicated in the
figure. The points with an input \mueff\ 
brighter than 22.0 are highlighted in blue.

As expected, the residuals in total magnitude anticorrelate with those in effective radius: 
where \gdd\ yields a higher than expected luminosity, it also yields a higher than expected effective radius.  This occurs for all S\'ersic indices.  
The oblique, green line shows the error correlation that would preserve the mean effective surface brightness within
one effective radius.  
The figure shows that \gdd\ introduces
a surface brightness bias: overluminous solutions have fainter $\langle \mu \rangle _{\rm e}$. 
The slope of this covariance is found to be S\'ersic index-dependent. The contraction or expansion
of the fitted functions with respect to the input parameters is not arbitrary, it depends
on the real profile of the underlying source being fitted. Although these observations merely confirm an expected 
behaviour given the experiment being run, they are important
because they allow us to use these simulations
to assess the errors on the output parameters. 

Comparable behaviour can be seen in the right panels of Fig.~\ref{fig:g2d_simulations_1}. If \gdd\ finds \Reff\ 
higher than the real value, the output $n$ is also higher than the
real value. This happens because $n$ is
essentially determined by the pixels 
in the object wings, which are much more abundant and have higher
weighting due to the lower Poisson noise, thus \gdd\ has to increase the model power in these wings. 

In addition, the fact that the S\'ersic index has such a large impact
on the behaviour of the residuals implies that this
parameter has to be included in our recipe for error assessment
(see also Marleau \& Simard 1998, their figure 11).

\begin{figure*}
\begin{center}
\includegraphics[scale=0.4]{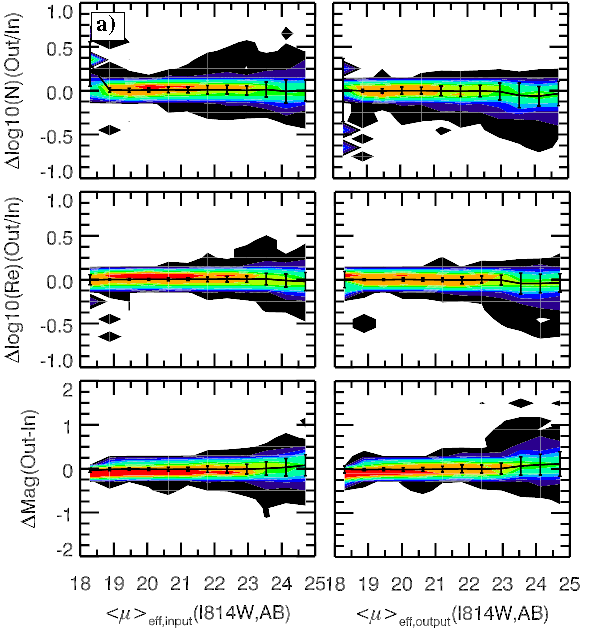}\includegraphics[scale=0.4]{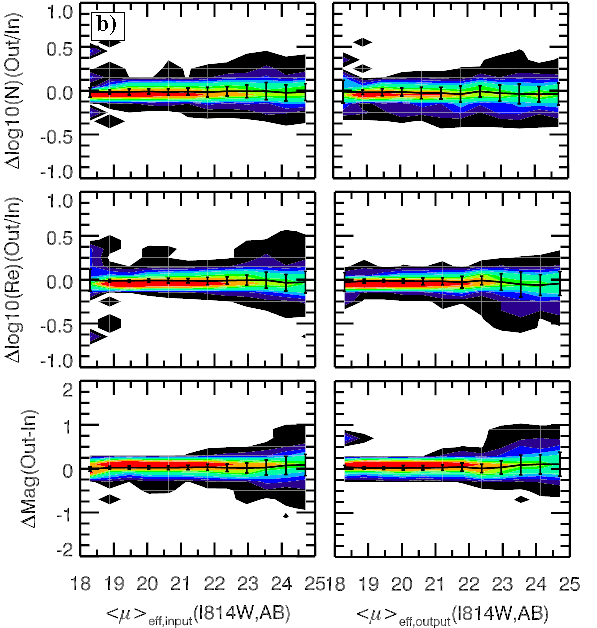}
\vspace{0.2cm}
\hspace{0.5mm}\includegraphics[scale=0.4]{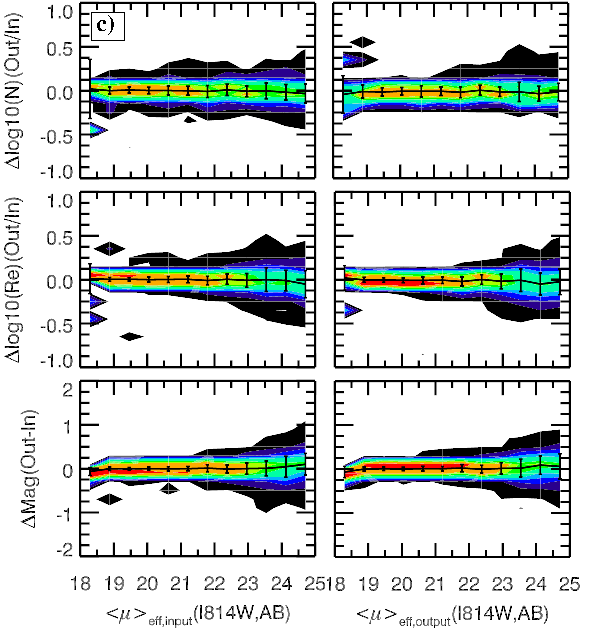}\includegraphics[scale=0.4]{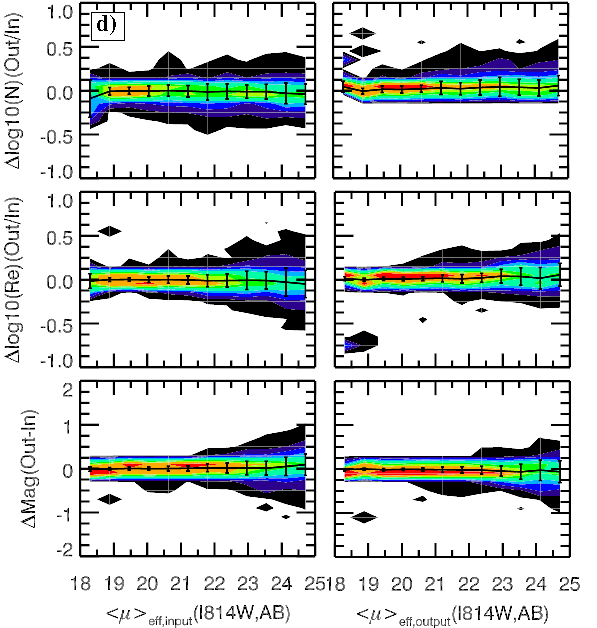}
\vspace{0.5cm}
\caption{
Magnitude, effective radius and S\'ersic index residuals for
  the \gdd\ simulations. This figure is divided into four quadrants,
  each with six panels. Top left quadrant (a) shows the residuals for
  simulations of galaxies with input $0.5 < n <  1.5$; top right quadrant (b)
  simulations with input $1.5 < n < 2.5$, bottom  left quadrant (c)
  simulations with input $2.5 < n < 3.5$, and bottom right quadrant (d)
  simulations with input $3.5 < n < 8$. In each quadrant, the top two
  panels show the residuals in S\'ersic index as a function of input \mueff\ (left) and 
  output \mueff\ (right) panels. The middle two panels show
the residuals in \Reff\ and the bottom two the residuals in
magnitude, again against input and output $\langle \mu \rangle _{\rm e}$. 
The lines with vertical error bars show the run of the median value of the residuals; the error 
bars are 1.5 times the inter-quartile width of the vertical distribution.
The colour coding shows the two dimensional histogram of the
density of the underlying points, normalised along the vertical
axis only. The lowest level (black) has a density $>$1\% of the maximum, and
the highest level (red) is $>$50\% of the maximum. Intermediate shades are
at 5\%, 10\%, 15\%, 20\%, 30\% and 40\%.\label{fig:g2d_simulations}}
\end{center}
\end{figure*}

\begin{figure*}
\begin{center}
\includegraphics[scale=0.4]{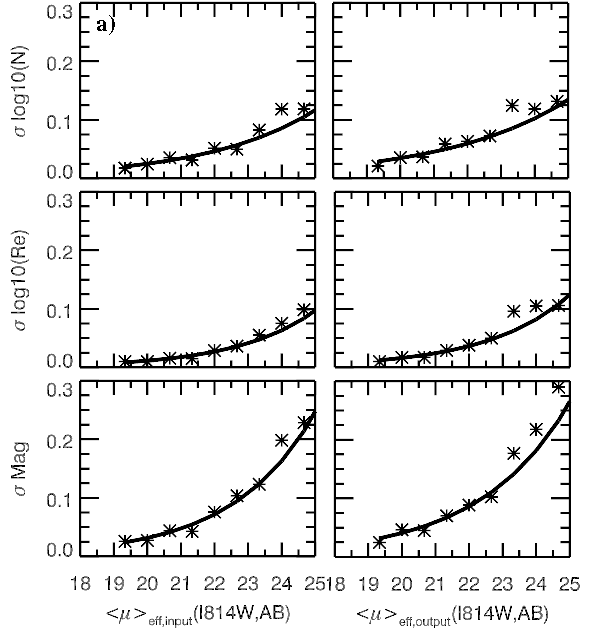}\includegraphics[scale=0.4]{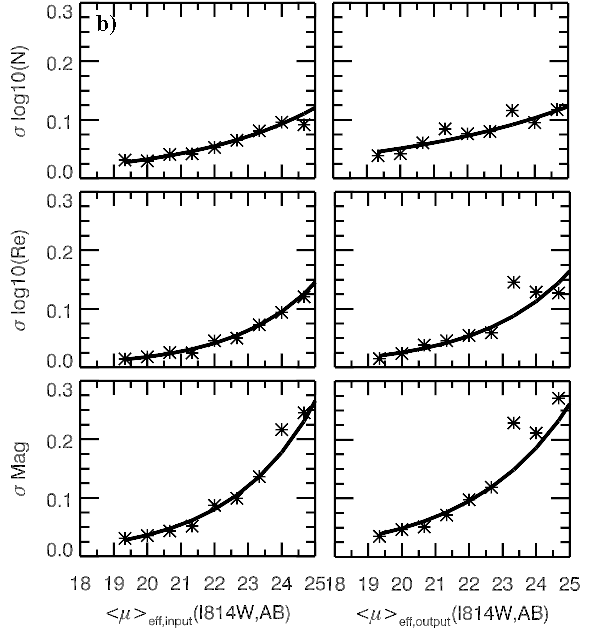}
\vspace{0.2cm}
\hspace{0.5mm}\includegraphics[scale=0.4]{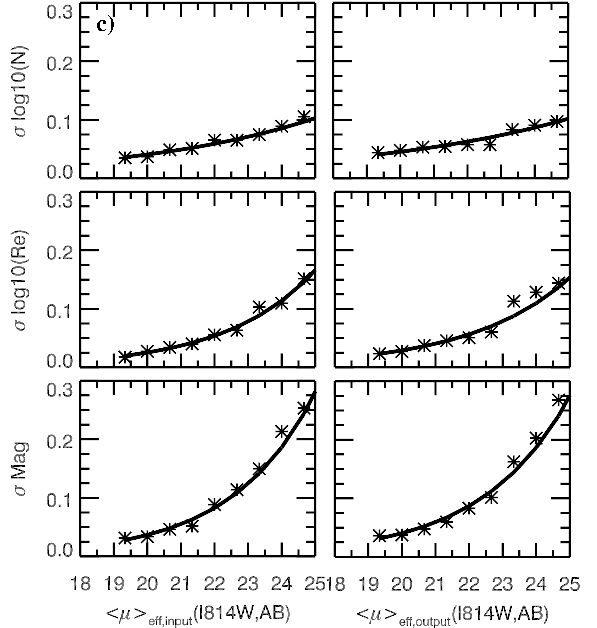}\includegraphics[scale=0.4]{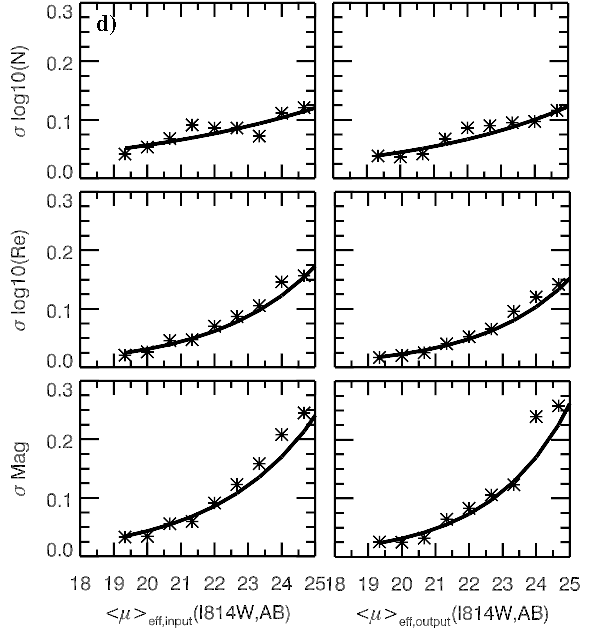}
\vspace{0.5cm}
\caption{
Amplitude of the error bars for the \gdd\ simulations plotted in Fig.~\ref{fig:g2d_simulations}.   
The figure layout is identical to that of Fig.~\ref{fig:g2d_simulations}.  
In each panel, the right panels provide a direct measure of the \mueff\ dependence 
of the non-systematic error in 
each of the model parameters $\Delta\log(n)$, $\Delta\log(\Reff)$ and $\Delta\mathrm{mag}$.
\label{fig:g2d_simulations_RMSerrors}}
\end{center}
\end{figure*}

Fig.~\ref{fig:g2d_simulations} shows the
magnitude, effective radus and S\'ersic index residuals against both input and output mean effective surface 
brightness for the 10000  simulations. This figure is divided
into four panels: for  S\'ersic indices of 
$0.5< n < 1.5$, $1.5 < n < 2.5$, $2.5 < n < 3.5$, and $3.5 < n < 8.0$, 
respectively. 
The black lines with error bars give the run of the median value of the residuals and a robust
estimate of the vertical $1\sigma$ scatter of the residuals around this latter median value.
These median values of the residuals and their associated  1$\sigma$ scatter have been 
calculated in equally spaced bins in \emph{input} (left panels) and \emph{output} (right panels)
mean effective surface brightness. 
For clarity, the vertical 1$\sigma$ scatter values are plotted separately in Fig.~\ref{fig:g2d_simulations_RMSerrors}. 

From the data presented in Figs.~\ref{fig:g2d_simulations} and~\ref{fig:g2d_simulations_RMSerrors} we find that,
for \mueff\ $\leq$ 24.0, the median values of $\Delta\log(n)$, $\Delta\log(\Reff)$ and $\Delta\mathrm{mag}$
(the systematic errors) are always less than 0.05, 0.04, and 0.08, respectively. These values are only
weakly dependent upon $n$, with the highest S\'ersic index bin having smaller systematic errors
(less than 0.05, 0.03, and 0.05 respectively). For brighter surface brightness (\mueff\ $\leq$ 21.0) these differences are
lower than 0.02, 0.03, and 0.04. The widths of the distributions, which we associate with the non-systematic error in
the recovery of the input values, increase toward lower surface brightness, and are, at the faintest limit,
0.12 dex, 0.15 dex, and 0.25 mag, for $\log(n)$, $\log(\Reff)$ and total magnitude, respectively. 
These values and Figures indicate that, with the use of the tailored masks, \gdd\ is indeed 
able to recover the input parameters accurately. The use of these customized masks and the use of the individualized 
search in the parameter space allows \gdd\ to have a better understanding of the galaxy flux and size. This naturally leads
to a better fit, free from the systematic errors that were detected
by H\"aussler et al.~(2007) and 
confirmed in Appendix~\ref{apx:g2d_gfit07} . 

Figs.~\ref{fig:g2d_simulations} and~\ref{fig:g2d_simulations_RMSerrors} indicate that it is reasonable to
use the output S\'ersic index and \mueff\ to derive realistic error estimates 
to the total magnitudes, effective radii and S\'ersic indices. 
Given the modest number of simulations,  we adopt a simple 
two-parameter approach based upon the output \mueff\ 
and  $n$. A single
straight line of the form:
\begin{equation}
 \log \sigma=\alpha \times
\langle \mu \rangle_{\mathrm{e,out}}+\beta,
\label{Eqn:errors}
\end{equation}
\noindent is fit to the robust 1$\sigma$ vertical
scatter around the median shown in the right panels of 
each quadrant of Fig.~\ref{fig:g2d_simulations_RMSerrors}. 
Although this functional form is expected for the
magnitude uncertainties, it is also used for the uncertainties in \Reff\ and S\'ersic
index for simplicity.
Table~\ref{Tab:g2d_simulations_table} shows 
the best-fit coefficients for these fits.

To assign meaningful and realistic statistical errors to any \gdd\
measurement, we first calculate the output {\mueff}. Next we evaluate the linear functional form given above
using the coefficients found in Table~\ref{Tab:g2d_simulations_table}.
The final uncertainty in the parameter of interest, as given in the structural catalogue, 
is the antilogarithm of the result.  

Finally, the run of the errors with \mueff\ and $n$, given in Fig.~\ref{fig:g2d_simulations_RMSerrors} 
and derived from the coefficients in Table~\ref{Tab:g2d_simulations_table}, 
allow us to infer a limiting output \mueff\
beyond which \gdd\ will not be able to successfully recover 
the true parameters. We choose an operational limit of 
an uncertainty of 0.25 mag. This limit corresponds 
to $S/N < 5.0$, and given the coefficients in the table, 
the corresponding limiting $\langle \mu \rangle _{\rm e}$
is 24.5 mag arcsec$^{-2}$. Since the magnitude is the first moment of the 
light distribution of any object, it will not be possible to 
reliably recover the remaining structural parameters, which would be
higher moments, from lower
surface brightness objects. 

\begin{table}
\begin{center}
\begin{tabular}{ll|rrr} \hline
Magnitude        & S\'ersic index		& $\beta$  & $\alpha$	      \\  \hline   
                 & $0.5<n<1.5$               		& -4.64 &  0.16 	      \\	   
                 & $1.5<n<2.5$               		& -4.22 &  0.15 	      \\	   
                 & $2.5<n<3.5$           		& -4.72 &  0.17 	      \\	   
                 & $3.5<n<8.0$	   			& -5.17 &  0.18 	      \\	   \hline
$\log {\Reff}(\mathrm{G2D/Model})$  & S\'ersic index    	& $\beta$  & $\alpha$         \\  \hline   
                 & $0.5<n<1.5$                      	& -5.34 &  0.18 	      \\	   
                 & $1.5<n<2.5$                      	& -4.83 &  0.16	      	      \\	   
                 & $2.5<n<3.5$                      	& -4.43 &  0.14 	      \\	   
                 & $3.5<n<8.0$          			& -4.93 &  0.16 	      \\	   \hline
$\log n(\mathrm{G2D/Model})$& S\'ersic index 		& $\beta$  & $\alpha$         \\ \hline	   
                 & $0.5<n<1.5$               		& -3.81 &  0.12	      	      \\	   
                 & $1.5<n<2.5$               		& -2.81 &  0.08 	      \\	   
                 & $2.5<n<3.5$               		& -2.74 &  0.07 	      \\	   
                 & $3.5<n<8.0$   					& -3.11 &  0.09 	      \\      \hline     
\end{tabular}
\caption{Table of coefficients required to use Equation~\ref{Eqn:errors} to
estimate the statistical errors on the total magnitude, \Reff\ and $n$, for
various ranges of output $n$, for the \gdd\ fits.} 
\label{Tab:g2d_simulations_table}
\end{center}
\end{table}

\section{Galfit fitting}
\label{sec:setup_galfit}

This section describes the \galfit\ (version 2.0.3c) setup used to fit
the program galaxies  
as well as the simulations carried out for error assessment. Nearly all galaxies 
included in the \sex\ catalogue presented in Paper II were fit, except for the 
sources that were originally buried in the extended haloes of large galaxies.
\galfit\ is capable of fitting multiple galaxies simultaneously.
Because its $\chi^2$ minimisation algorithm is based on a
gradient method it is  significantly faster than {\gdd}. However the algorithm
is susceptible to getting stuck in a local minimum.
For the fit to converge quickly to the correct values
it is essential that the initial values of the parameters are as close
to the real solution as possible.

Most initial parameters that we use for fitting are based on the
\sex\ catalogue. When fitting large numbers of galaxies any manual
intervention is extremely time consuming. Therefore, we decided to
make use of \aw\footnote{www.astro-wise.org}, which provides a facility 
for the structural analysis of large datasets.

\subsection{\galfit\ setup in \aw}

\aw\ is an information system and
environment for large imaging datasets, up to the Petabyte regime, with multiple 
users around Europe. In \aw\ one can archive raw
data, calibrate data and perform scientific analysis storing all
results. Valentijn et al.~(2007) provide a technical description of the information system, and Sikkema (2009) describes the data reduction pipeline.

The scientific analysis components in \aw\ include, among others,
routines for source extraction (using \textsc{SExtractor}),
variability analysis, photometric redshifts and  galaxy
surface photometry fitting using {\galfit}. The \galfit\
implementation within \aw\  automatically produces the specified 
postage stamps of the sources, runs \galfit\ itself and stores the
configuration and results for all sources in a
database. \aw\ enables full backward-chaining of data lineage in
general. This means that model image,
residual image and customized inspection plots can be created 
anytime upon request.

We use the \galfit\ setup of H\"aussler et al.~(2007) as a starting
point for our fitting scheme. Here we present a detailed description 
of the fitting process.

First, \aw\ makes a postage stamp of each source.  The size chosen is
slightly larger than that of  H\"aussler et al., and is determined
from the \sex\ image size measurements, such that 
\begin{equation}
\mathtt{size} = 4 \cdot \mathtt{A}\_\mathtt{IMAGE} \cdot \mathtt{KRON\_RADIUS}. 
\end{equation}
\noindent Next, a sigma image is created from the Inverse Variance Map. This
sigma image is modified to take into account Poisson noise from the 
sources as well.

Nearby sources are masked according to the \sex\ segmentation image. 
In line with the GEMS results and encouraged by the results with \gdd\
we expand the masks for nearby sources by using elliptical apertures 
with semi-major axis $4\times${\verb1A_IMAGE1} and with ellipticity
and position angle as determined by \textsc{SExtractor}. Sources for which this mask 
overlaps with the mask of the main source are fitted by \galfit\
together with the main source.  

The implementation of the S\'ersic profile in \galfit\ has 8 free
parameters. Of these, we leave the diskiness/boxiness
parameter fixed so that all isophotes describe perfect ellipses. All 
other parameters are left free. In addition to this, we leave the sky 
free, although we do not allow for any gradient in the sky. We do, 
however, constrain $n$ to the interval
$[0.5,8.0]$. Gradient based fitting methods do require an initial
guess for all parameters. Except for the S\'ersic index, which we
initialize as $n = 1.5$ for all sources, initial guesses for each source
are based on parameters from the \sex\ catalogue: for \Reff\ we use 
\texttt{FLUX\_RADIUS[3]}, for total magnitude we use
\texttt{MAG\_ISO}. The axis ratio and position angle are initialized
from \texttt{ELLIPTICITY} and \texttt{THETA\_IMAGE}.

\aw\ uses these parameters to write a configuration file for
{\galfit}. In the case of mulitobject fitting, we determine the input 
parameters for the secondary objects in the same way, with the 
exception that we keep the position of the source fixed if its centre 
is outside the postage stamp.

\subsection{Shot-noise Simulations}
\label{galfit:simulations}

Similarly to what was done for \gdd\ (Sect.~\ref{g2d:simulations}), 
an extended set of simulations were performed.   
The simulations serve three main purposes. 
First, they allow us to test our \galfit\ setup, by identifying
biases in the fits.  
Second, they allow us to infer realistic errors of the output parameters. 
Like \gdd\ and other fitting codes, \galfit\ 
tends to underestimate errors on the
fitted parameters (cf. H\"aussler et al.~2007). Our simulations allow us to assign errors to fitted 
parameters which are more realistic than the standard \galfit\ errors.
(Our errors are still lower limits because images of real galaxies 
deviate from the perfect S\'ersic model with concentric, coaxial isophotes.)
Finally, the simulations allow us to define limits for the minimum 
signal-to-noise required for reasonable fits. The simulations were
not designed to test the performance of \galfit\ in crowded areas, 
which has already been extensively discussed by H\"aussler et
al.~(2007).

\galfit\ is wrapped in \aw\ using the \textsl{python} 
language which allows for straightforward customization and script 
writing for a specific science case. We adapted the {\rm python} 
code in \aw\ to create simulated galaxies, insert them into images,
create source lists and then to run \galfit\ on them.
As is the case with \gdd, \galfit's ability to correctly  fit a given galaxy varies with the
intrinsic parameters of the fitted galaxy. Hence, to assign errors to
the fit parameters of real galaxies we require the results of a large
number of simulated galaxies with similar output parameters. 

In our approach we created a mock catalogue of 200,000 galaxies. 
The parameter ranges used are listed in Table~\ref{tab:galfit_sim_pars}.
Each parameter samples the given range, either uniformly or uniformly in 
the log as indicated in the Table. 
The parameter ranges were chosen so that the distributions of output parameters 
bracket the distributions found in the data.  
When generating these parameters, we
avoided the edges of the frame and applied a hard cut-off in 
$\mu_e$ to avoid any detection problems with {\sex}. 

After each model galaxy had been fitted by \galfit, the distributions of 
the differences output \textit{minus} input
were binned in order to determine the variation of the errors with key 
\textit{output} parameters such as magnitude, surface brightness and S\'ersic index.  
We chose six bins in magnitude, ten bins in log(\Reff), two bins in
ellipticity and five bins in $n$ (600 bins in total). To
minimise the uncertainty on the errors, one would like to have as many 
galaxies per bin as possible. Our 200,000 models yield $\sim$330
per bin, so that, even though for some galaxies the output bin will
be different from the input bin and certain output bins will be more
sparsely populated than others, the relative uncertainty on the errors 
(assuming they are Gaussian) is always less than 10\%.

\begin{table}
\begin{center}
\begin{tabular}{lll}
\hline
\hline
parameter & range & $\#$bins\\
\hline
x (pix) & 400.0\ldots4000.0 & -\\
y (pix) & 400.0\ldots4000.0 & -\\
mag & 20.0\ldots25.0 & 6\\
\Reff & 2.0\ldots60.0 (log) & 10 \\
n & 0.5\ldots6.0 (log) &	5\\
$\mu_e$ & $<$25.5 & -\\
ell & 0\ldots0.8 & 2\\
pos & 0.0\ldots180.0 & -\\  \hline
\end{tabular}
\end{center}
\caption{Input simulation parameters of \galfit\ single Sers\'ic galaxies. 
All quantities are cast uniformly in the given range, 
except \Reff\ and $n$, which are cast uniformly 
in logarithmic space. The number of bins denotes into how many bins 
the range was divided for the final error assesment. In the \galfit\ simulations the
effective surface brightness rather than the mean effective surface brightness was used to define the 
bins.}\label{tab:galfit_sim_pars}
\end{table} 

\galfit\ itself was used to generate 
the artificial galaxy models. Although one might argue that using 
\galfit\ to make the two-dimensional images to which itself it should fit 
models is 
doubtful, we stress that \galfit\ has been tested extensively and that, in our
opinion, 
it is doing at least a better job than \textsc{IRAF.artdata}, which
does not sufficiently oversample in the centres of galaxies. Our
simulation setup takes into account convolution of the model 
galaxies with a \dt\ PSF. Before injecting them into real ACS
observations, Poisson noise was added to these galaxies. 
To avoid any crowding, we used only 100 models per ACS-frame, so that we
ended up with 2000 frames, each with 100 artificial galaxies on top of
the $\sim$2500 sources already present. Simulated galaxies were
injected into Visit 90, because this frame is relatively empty and on does not
suffer from any missing dithers.

On each frame with simulated galaxies we ran \sex\ using the same 
configuration as was used for the real data (see Paper II). We associated 
our list of simulated sources with the sources detected by \sex\ by 
demanding that they be at most 14 pixels away from the closest source 
in the catalogue. A 
small fraction ($\sim$1\%) of sources were not detected by {\sex}. In a
small number of cases, \sex\ can be confused by proximity to or even
blending with a source already present in the frame.

\begin{figure*}
\begin{center}
\includegraphics[scale=0.4]{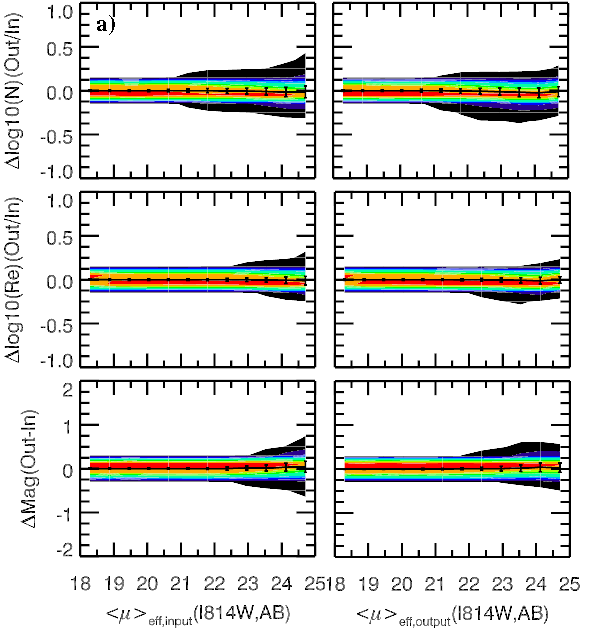}\includegraphics[scale=0.4]{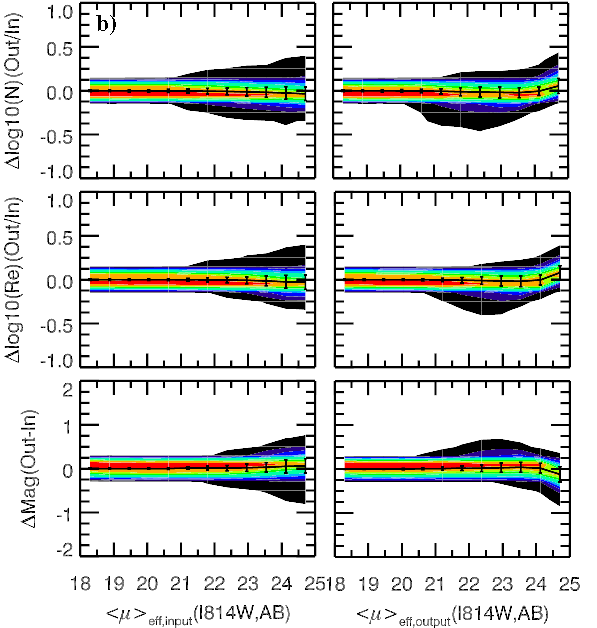}
\vspace{0.2cm}
\hspace{0.5mm}\includegraphics[scale=0.4]{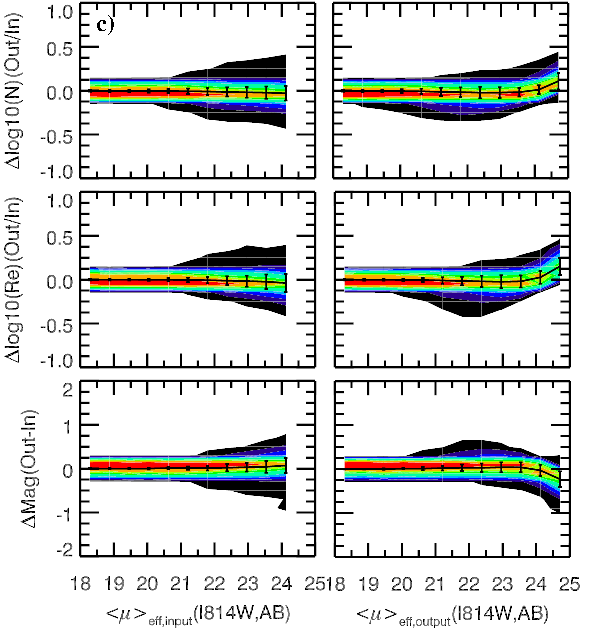}\includegraphics[scale=0.4]{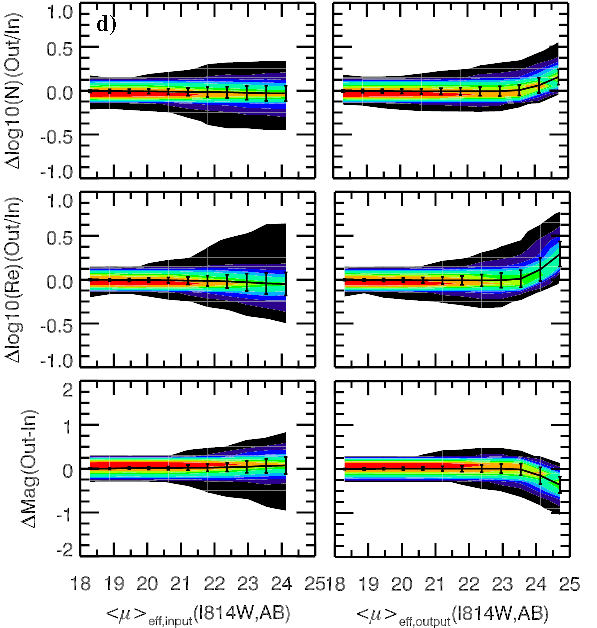}
\caption{
Magnitude, effective radius and S\'ersic index residuals for
  the \galfit\ simulations. This figure is organized as Fig.~\ref{fig:g2d_simulations}. 
  Magnitude, effective radius and S\'ersic index residuals for
  the \gdd\ simulations. This figure is divided into four quadrants,
  each with six panels. 
  Top left quadrant (a) shows the residuals for
  simulations of galaxies with input $0.5 < n < 1.5$; top right quadrant (b)
  simulations with input $1.5 < n < 2.5$, bottom  left quadrant (c)
  simulations with input $2.5 < n < 3.5$, and bottom right quadrant (d)
  simulations with input $3.5 < n < 6$. In each quadrant, the top two
  panels show the residuals in S\'ersic index as a function of input \mueff\ (left) and 
  output \mueff\ (right) panels). The middle two panels show
the residuals in \Reff\ and the bottom two the residuals in
magnitude, again against input and output $\langle \mu \rangle _{\rm e}$. The lines 
with vertical error bars show the run of the median value of
the residuals. As in Figure~\ref{fig:g2d_simulations} the error bars are given
by 1.5 times the interquartile range.
The colour coding shows the two dimensional histogram of the
density of the underlying points, normalised along the vertical
axis only. The lowest level (black) has a density $>$1\% of the maximum, and
the highest level (red) is $>$50\% of the maximum. Intermediate shades are
at 5\%, 10\%, 15\%, 20\%, 30\% and 40\%.\label{fig:galfit_simulations}
}
\end{center}
\end{figure*}

\begin{figure*}
\begin{center}
\includegraphics[scale=0.4]{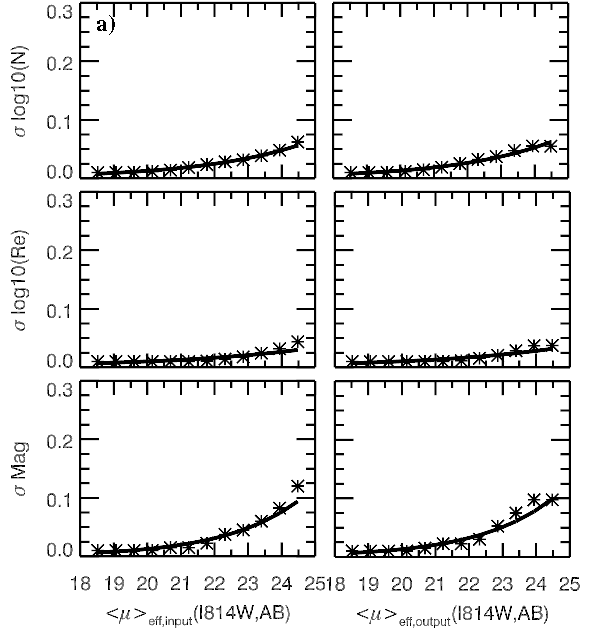}\includegraphics[scale=0.4]{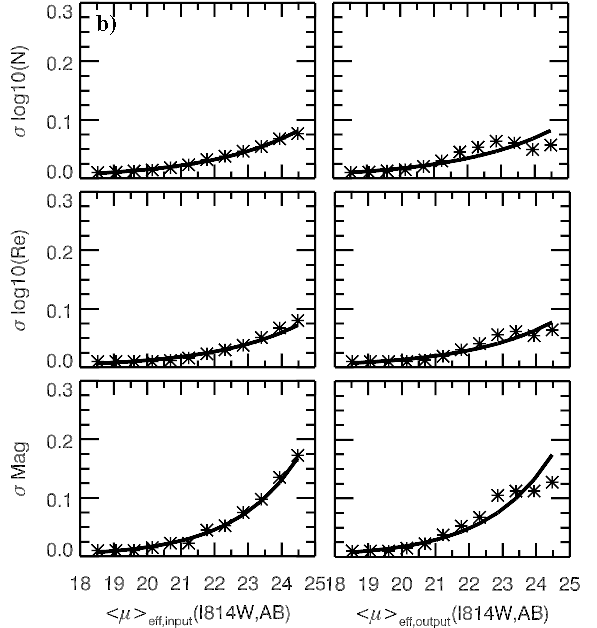}
\vspace{0.2cm}
\hspace{0.5mm}\includegraphics[scale=0.4]{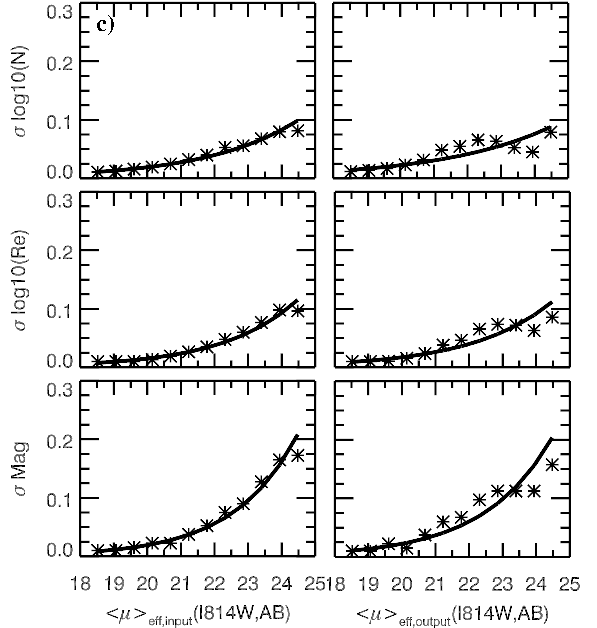}\includegraphics[scale=0.4]{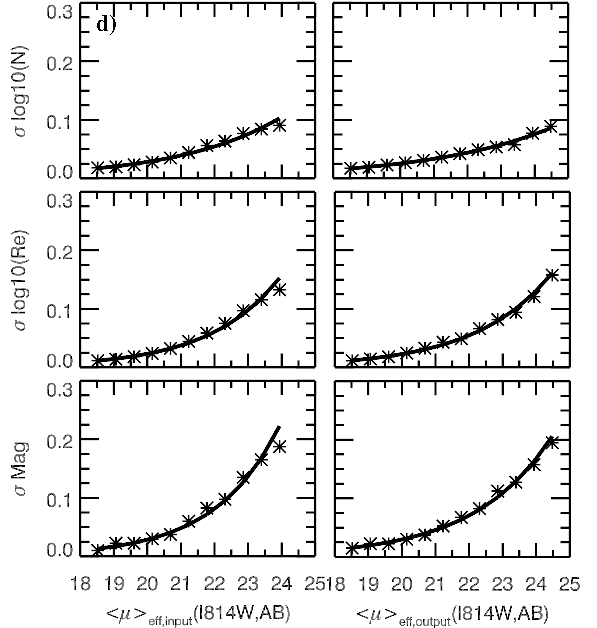}
\vspace{0.5cm}
\caption{
Amplitude of the error bars for the \galfit\ simulations plotted in Fig.~\ref{fig:galfit_simulations}.   
The figure layout is identical to that of Fig.~\ref{fig:galfit_simulations}.  
In each panel, the right panels provide a direct measure of the \mueff\ dependence 
of the non-systematic error in 
each of the model parameters $\Delta\log(n)$, $\Delta\log(\Reff)$ and $\Delta\mathrm{mag}$.
\label{fig:galfit_simulations_RMSerrors}
}
\end{center}
\end{figure*}

Results of the simulations are shown in
Figs.~\ref{fig:galfit_simulations} and \ref{fig:galfit_simulations_RMSerrors}.  
As in Fig.~\ref{fig:g2d_simulations}, Fig.~\ref{fig:galfit_simulations} shows residuals of 
the logarithm of the S\'ersic index $\log(n)$, 
the logarithm of the effective radius $\log(\Reff)$, and the total magnitude, 
against input (left) and output (right) mean effective surface brightness \mueff.
In general, the distributions of the output parameters around the mean
have non-Gaussian, extended wings. Hence, a standard RMS error 
does not allow for a straightforward interpretation. Rather than
determining the RMS of a outlier-clipped sample, we use a 95\% 
confidence interval determined from the interquartile range per
surface brightness or magnitude bin. The symmetrized intervals are
used as errorbars in the plots in Fig.~\ref{fig:galfit_simulations}, 
and plotted again in Fig.~\ref{fig:galfit_simulations_RMSerrors}. 

The results are excellent, and, overall, similar to those obtained with \gdd\ (Sect~\ref{g2d:simulations}). 
Up to $\mueff = 24.0$, 
the median differences, or systematic errors, are below 0.04 dex, 0.02 dex, and 0.04 mag, in 
$\log(n)$, $\log(\Reff)$ and total magnitude, respectively, except for the highest S\'ersic index bin  
where the differences at $\mueff = 24$, are always lower than 0.06 dex, 0.06 dex or 0.07 mag. 
The slight bias pattern that appears for $\mueff > 24.0$ and high S\'ersic indices, (Fig.~\ref{fig:galfit_simulations}c,d), is an boundary effect of the simulation setup.  
Output models are brighter, and have larger \Reff, 
than the input values due to the fact that input models only reach 
$\mueff\ \lesssim 24.0$. 
The region with output $\mueff > 24.0$ is only populated with models for which \galfit\ 
has found a solution with fainter \mueff.  Because 
the errors in \Reff\ and \mueff\ are coupled, these models must have positive \Reff\ residuals,
as observed in the middle-right panels of Fig.~\ref{fig:galfit_simulations}c,d).

The run of non-systematic errors with input and output mean effective surface brightness
(Fig.~\ref{fig:galfit_simulations_RMSerrors}) 
shows similar behaviour to those of the \gdd\ errors (Fig.~\ref{fig:g2d_simulations_RMSerrors}). 
For a given surface brightness, \galfit\ errors tend to be slightly smaller than \gdd\ errors but the 
differences are not meaningful, given that \gdd\ values are more uncertain owing to the lower 
number of \gdd\ simulations.

In Table~\ref{Tab:galfit_simulations_table} we present the parameters needed to estimate the uncertainties 
of the GALFIT output parameters, as was previously done in Table~\ref{Tab:g2d_simulations_table} for GIM2D. 
This gives the coefficients for the fits of Equation~\ref{Eqn:errors}. The best-fit relations were then applied 
to extract the error estimates for each of the S\'ersic parameters 
that are given in the published structural catalogue.  

The simulations were also used to provide a reasonable faint limit in surface brightness 
that gives realistic results.  A conservative approach was used to define this limit, 
which we also set to be the same as that applied for the \gdd\ fits.

\begin{table}
\begin{center}
\begin{tabular}{ll|rrr} \hline
Magnitude        & S\'ersic index & $\beta$  & $\alpha$  \\ \hline
                & $0.5<n<1.5$               & -1.70    & 0.036     \\                    
                & $1.5<n<2.5$               & -1.30    & 0.020     \\                   
                & $2.5<n<3.5$               & -1.65    & 0.035      \\                   
                & $3.5<n<8.0$			   & -2.62    & 0.079      \\    \hline
$\log \Reff(\mathrm{GF/Model})$  & S\'ersic index    & $\beta$ & $\alpha$  \\ \hline
                & $0.5<n<1.5$                      & -3.90   & 0.12     \\                
                & $1.5<n<2.5$                      & -3.57   & 0.11     \\             
                & $2.5<n<3.5$                      & -4.42   & 0.15     \\                   
                & $3.5<n<8.0$          & -3.58   & 0.11     \\    \hline
$\log n(\mathrm{GF/Model})$& S\'ersic index & $\beta$ & $\alpha$  \\ \hline
                & $0.5<n<1.5$               & -4.09   & 0.13     \\                
                & $1.5<n<2.5$               & -3.17   & 0.092     \\                    
                & $2.5<n<3.5$               & -3.01   & 0.084      \\                     
                & $3.5<n<8.0$   & -2.77   & 0.075      \\         \hline
\end{tabular}
\caption{Table of coefficients required to use Equation~\ref{Eqn:errors} to
estimate the statistical errors on the total magnitude, \Reff\ and $n$, for
various ranges of output $n$, for the \galfit\ fits.}
\label{Tab:galfit_simulations_table}
\end{center}
\end{table}

\begin{table}
\begin{center}
\begin{tabular}{lcc}
\hline
\hline
Parameter & Range & Remark\\
\hline
x & 400.0 \ldots 4000.0 & \\
y & 400.0 \ldots 4000.0 &\\
mag & 22.0 \ldots 26.5 & \\
\Reff & 0.1\ldots5.0 & Log\\
n  & 0.5\ldots6.0 & Log\\
$\mu_e$ & $<$25.5 & \\
ell & 0\ldots 0.8 & \\
pos & 0.0\ldots 180.0 & \\  \hline
\end{tabular}
\end{center}
\caption{Parameters of simulated galaxies for the small radii simulations.
Same comments as for Table~\ref{tab:galfit_sim_pars} apply. $R_{\mathrm{eff}}$ and $n$ are
logarithmically spaced. As with Table~\ref{tab:galfit_sim_pars} the limit in surface brightness is defined in
terms of $\mu_e$ not {\mueff}}
\label{tab:galfit_sim_pars_small_radii}
\end{table}

\begin{figure}
\includegraphics[scale=0.45,angle=0]{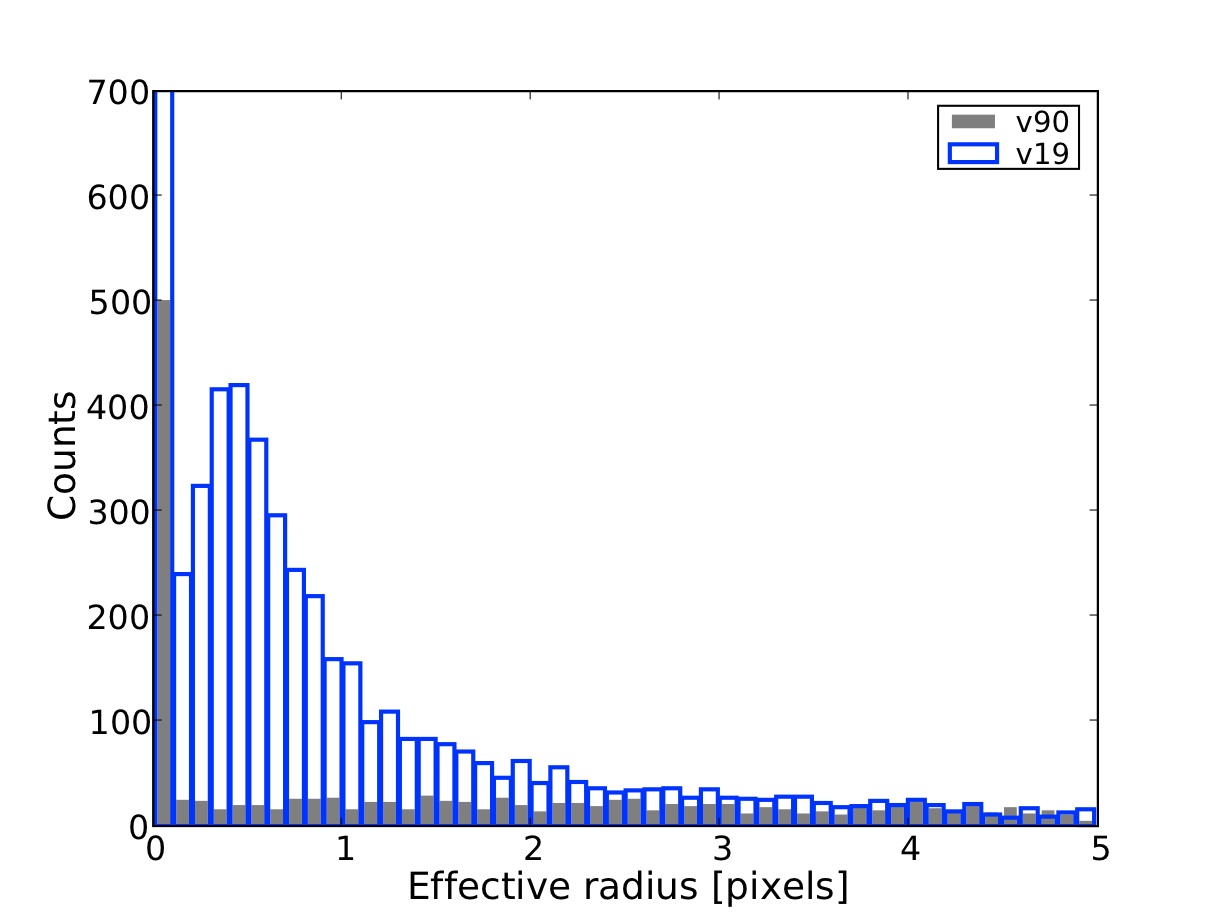}
\caption{
Histogram of Galfit Measurements of the \Reff\ for the sources found in visits 19 and 90. The filled histogram represents the sources in visit 90, 
i.e. a field with few Coma galaxies. The open histogram is the histogram of visit 19, with many Coma galaxies and globular clusters. }
\label{fig:sim_logre_vs_mu}
\end{figure}

\subsection{Small-size limit of GALFIT}
\label{cut:re}

With the discovery of 16 new Ultra Compact Dwarfs in the
Coma Cluster (Chiboucas et al.~2009) it is important to quantify how
small an effective radius we can measure.
To see how well \galfit\ can 
recover radii and magnitudes for small sources, 20000 further simulations were carried out. The parameter 
space covered by this new set of simulations is presented in 
Table ~\ref{tab:galfit_sim_pars_small_radii}.

We find from this new set of simulations that the recovered effective radii
are unbiased,  however for very small sources ($\Reff < 0.5$ pixels) \galfit\ sometimes falls back to a 
hard-coded lower limit of $\Reff = 0.01$ pix. This means that even for the perfect conditions assumed in the simulations, the number of 
recovered sources that have \Reff\ around 0.5 pix will be fairly
incomplete. It is very difficult for \galfit\ to differentiate between
a genuine point source and a small, yet extended, $\Reff < 0.5 $ pixel source.
These simulations assume perfect knowledge about the PSF, and model only the
background contribution to the noise.
Furthermore, the sources were injected on a relatively empty frame (visit 90). To see how 
well \galfit\ performs on real point sources, we inspect the effective radii of sources in visit 19, the visit 
which covers  the galaxy NGC4874. A large number of the sources in this visit are
thought to be globular 
clusters (Peng et al.~2010), which should be unresolved at the distance of 100 Mpc. Fig.~\ref{fig:sim_logre_vs_mu} 
shows a histogram of \emph{measured} effective radii for this visit. 
It looks as if the distribution of sources is shown by a powerlaw, plus an 
additional group of sources distributed around $\Reff \sim 0.5$ pix, 
with a standard deviation $\sim 0.3$ pix. We conclude that \galfit\ output giving 
$\Reff \sim 0.5$ pix is likely to come from point sources.  From the shape of the 
blue histogram in Fig.~\ref{fig:sim_logre_vs_mu} we infer that $\Reff > 1$ pixel provides a 
robust lower limit for \galfit\ \Reff's.

\section{Goodness of fit indices}
\label{sec:goodnessoffit}

Although the S\'ersic model is a well-known and tested 
fitting function, real galaxies are more complicated. 
They present, among many other features, 
stellar bulges, star forming regions, AGNs, spiral 
arms, extended haloes, and central star clusters.

In this section we define two complementary diagnostic indices, each designed to
address in a different way the question of whether the S\'ersic model
is an adequate fit, given the available data, or whether a more
complicated function or extra components are required.  
These diagnostics are calculated for all fits.

Following Blakeslee et al. (2006) we define the \textit{Residual Flux Fraction} (RFF) as:
\begin{equation}
\mathrm{RFF}=\frac{\left(\sum_i{\mathrm{|Res_i|}}-0.8\times\sigma_{\mathrm{image}}\right)}{\mathrm{FLUX\_ISO}},
\end{equation}
\noindent
where the summation is over all pixels within the \texttt{ISOAREA} of that particular object, $|\mathrm{Res}_i|$ is the absolute 
value of residual image obtained by subtracting the best-fit model
from the real galaxy image, and \texttt{FLUX\_ISO} is the total flux within 
the \texttt{ISOAREA}, which can be taken from \textsc{SExtractor}.
\indent

This index measures that part of the sum of the modulus of the pixels in the residual image which 
cannot be explained by the experimental error. The image variance is
obtained from the usual CCD equation as:
\begin{equation}
\sigma_{\mathrm{image}}^2 = {\sigma_{Bkg}^2 +\frac{S}{g}},
\end{equation}
\noindent
where $S$ is the value of the model for that pixel, and $g$ is the effective gain, which can be found from the \sex\ parameters
for the particular image.
\indent

RFF quantifies the residual signal which cannot be explained by arguing that
the fitting codes  found a suboptimal minimum. It is best understood as a 
hypothesis testing procedure.

If the real galaxy had  a pure S\'ersic profile, 
both \gdd\ and \galfit\ could find a model providing an exact fit to the galaxy.
However, even in this optimal case, the errors associated with the
readout noise and photon shot noise imply that the residual image will not be blank.
In the case of independent errors, the properties
of the residual image would be very similar to those of  gaussian white noise, with a spatially varying $\sigma$.
The expectation value of the sum of the absolute values of these
residuals is $\sum{0.8\times\sigma_{\mathrm{image}}}$. Therefore, the expectation value of the numerator
of the RFF is 0.0, \emph{should galaxies be pure S\'ersic models}. Since the denominator is a normalization factor, 
the expected value of the RFF itself is 0.0 for such a model.
Any positive or negative deviation from a pure S\'ersic model will increase the RFF.
From our own visual experience with this index, we find that fits with
a RFF larger than 0.11 indicate that a more complex fit is required. This number was
agreed after independent experiments made by CH and RG.

The RFF diagnostic does not work well for objects with large
\texttt{ISOAREAS}, and low $\langle \mu \rangle _{\rm e}$. In these cases, as both the galaxy and the model decay 
towards zero at large radii, the outer areas will dominate the RFF calculation, and 
even though the fit may have complicated
and highly non-gaussian residuals at small radii, RFF will still be small.

Hence we require a second diagnostic, which should be particularly sensitive to the central
residuals. This is calculated from the same set of pixels within the isophotal area
of the target galaxies as RFF.

This complementary parameter is the \textit{Excess Variance Index} (EVI). It is defined as:
\begin{equation}
\mathrm{EVI}=\frac{1}{3}\times\left( \frac{\sigma_{R}^{2}}{<\sigma_{\mathrm{image}}^{2}>}-1\right),
\end{equation}
\noindent
where $\sigma_R$ is the root mean square of the residuals within the \texttt{ISOAREA}, and $<\sigma_\mathrm{image}^2>$ is
the mean value of the square of the noise within the same area. EVI gives a measure of the granularity excess
of the actual residuals with respect to the expected granularity. 
\indent

For pure S\'ersic models, the numerator in the EVI expression above would simply be a sum of squares
of normally distributed random variables with zero mean and pixel-dependent variance. The expectation
value of the numerator would be, in this case, the value of the denominator. Hence, the parenthesized expression
in the EVI definition has an expectation value of 0.0, and strong deviations from 0.0 indicate that
the underlying real galaxy differs significantly from a S\'ersic model. Given the EVI definition, such deviations
would most likely be caused by the points with the largest intrinsic standard
deviations (i.e., the inner points). This makes this index particularly sensitive to the
structure of the residuals at small radii. The $1/3$ factor in the definition of the EVI
was later added so that residual images with an EVI larger than approximately 1.0 present complicated
substructure in the inner parts of the fits, whilst objects with an EVI smaller than approximately 1.0
show acceptable fits. In the end, after several iterations of this by  eye calibration carried out by CH and RG, it was 
concluded that residuals with an EVI larger than 0.95 indicate complex residuals.

\subsection{Comparison of the diagnostic indices}

\begin{figure}
\begin{center}
\includegraphics[width=8.5cm]{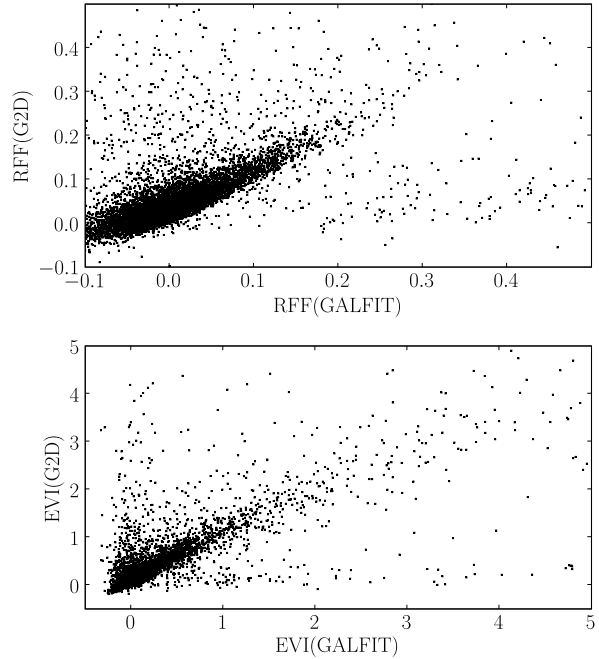}
\caption{Top panel: Residual Flux Fraction from \textsc{Galfit} plotted against that from \textsc{Gim2d};
  Lower panel: Excess Variance Index from \textsc{Galfit} plotted against that from \textsc{Gim2d}.}
\label{r1r2comp}
\end{center}
\end{figure}

Since the RFF and EVI indices were designed to see whether a single-S\'ersic fit
is sufficient, it is reasonable to ask how well the \gdd\ and \galfit\ agree on
this issue. As a first check, we plot in Fig.~\ref{r1r2comp} the RFF and EVI
indices of \galfit\ against their \gdd\ counterparts. There is a strong
correlation between the indices for the two codes. However, there is
also a small offset present, in the sense that the \gdd\ indices are
on average slightly higher than the \galfit\ indices. We attribute this
to a different treatment of the noise in the calculation of the two
indices (the \galfit\ indices were calculated using the sky noise from
the IVM maps, whereas for the \gdd\ indices we used the noise as estimated by \gdd\ itself.)  
For high values of RFF and EVI the correlation breaks down. 
Especially for EVI, there are galaxies where the \gdd\ index is normal,
but the \galfit\ index is larger. In part this is the result of fitting
low S/N sources, where \galfit\ has a preference for fitting a low
surface brightness model to fluctuations in the background instead of fitting the source itself. 

\begin{figure}
\begin{center}
\includegraphics[scale=0.4]{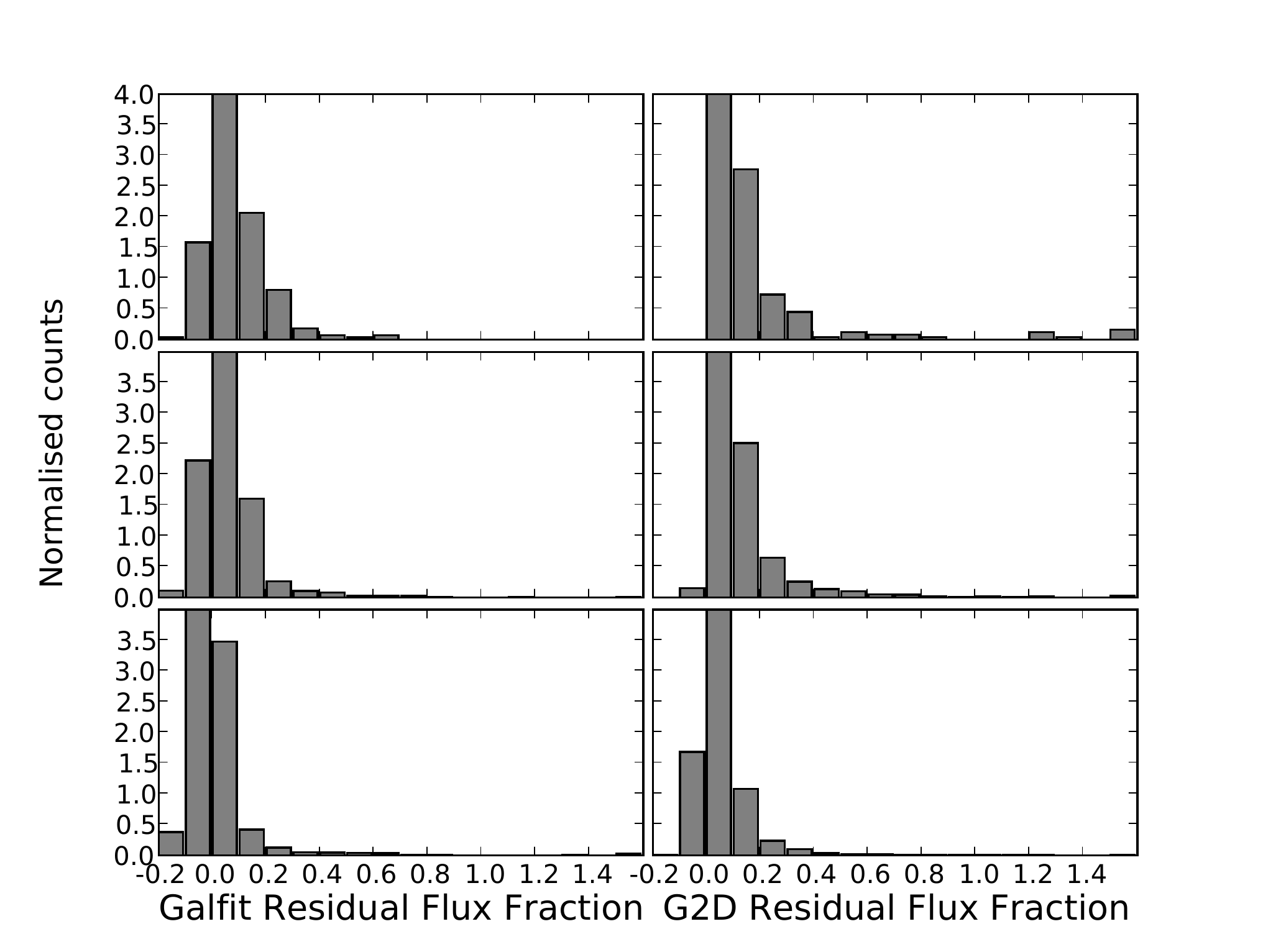}
\caption{The Residual Flux Fraction calculated for our magnitude,
  size, and surface brightness clipped sample on the residual images
  of both codes. The sample is divided in three magnitude bins:
  17.5-19.0 in the upper panel, 19.0-22.0 in the middle panel and
  22.0-25.0 in the lower panel.
}
\label{gf_g2d_rff}
\end{center}
\end{figure}

In Fig.~\ref{gf_g2d_rff} we show histograms of the Residual Flux
Fraction for different magnitude bins. If all sources were perfect
S\'ersic galaxies, these should take the form of a Gaussian, where the 
width is dependent on the signal-to-noise.  We find many faint sources
with RFF $<$ 0. We interpret this as overfitting: RFF $<$ 0 means that
there is less noise in the image than expected, so, the code has
modelled the noise away. For larger galaxies this problem is of course
less severe, as the code has to model more pixels with higher signal-to-noise, so it has less freedom.

\begin{figure}
\begin{center}
\includegraphics[scale=0.4]{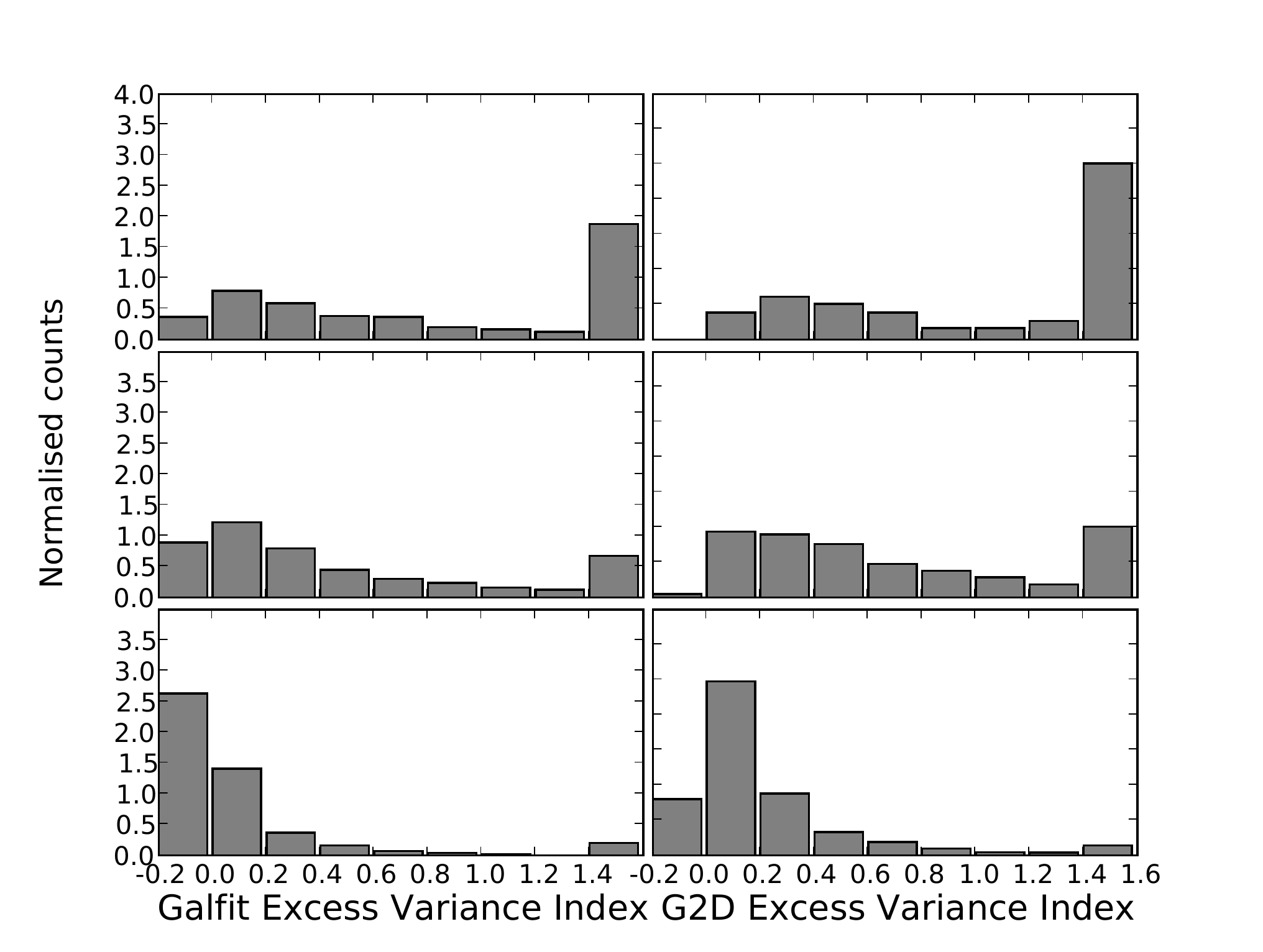}
\caption{The Excess Variance Index, see caption of Fig.~\ref{gf_g2d_rff} for more explanation.}
\label{gf_g2d_evi}
\end{center}
\end{figure}

Fig.~\ref{gf_g2d_evi} shows the calculated Excess Variance Indices,
for the same magnitude bins. Although the faint sources centre  around
zero, the brightest magnitude bins show a large fraction of high
EVIs. It is only in the brightest galaxies that we can see the
deviation from the S\'ersic profile (top panels), because of the high signal-to-noise
and spatial resolution. For the faint sources, the data
are not good enough to detect any granularity.

These indicators will allow separation into subsamples of galaxies which are 
well represented by the S\'ersic function, for studies of the scaling
relations or for separation by morphological
class, or of galaxies for which additional components, truncation or extension of
the light distributions show that they must be fitted with more complex functions.

\section{The Catalogue}
\label{sec:catalogue}

This section introduces the main result of the present paper: the catalogue of structural parameters 
for the Coma-ACS survey based on single S\'ersic fits.  
We provide a description of the catalogue and its limits in magnitude, radius, 
and effective surface brightness.  

The parameters presented in the final catalogue are summarized in Table~\ref{tab:cat_explanation}, which
presents the table headings of the released source lists together with a brief explanation of their meaning.
In addition to best-fit values for both \gdd\ and \galfit\ fitting, the catalogue includes coordinates and 
\sex\ photometry, and goodness of fit \texttt{RFF} and \texttt{EVI} for both codes.  

Recalling that all 74992 galaxies from the \sex\ photometric catalogue from Paper~II were fitted with 
\gdd\ and \galfit,  
the full structural catalogue presented here contains all galaxies from the complete photometric catalogue 
which fulfill the selection criteria described in Sect.~\ref{cut:mag}.  This amounts to 8814 sources.  
The full version of the structural parameter catalogue can be found in the electronic version of this 
paper, in the \aw\ website and  the HST MAST 
archive.

\begin{table}
\begin{tabular}{|l|l|p{16em}|}
\hline 
Col.&Parameter& Meaning. \\
\hline
1&\texttt{COMA\_ID}         & Name of source, as it appears in Paper II. \\
2&\texttt{RA (J2000)}               & Right ascension of source. \\
3&\texttt{DEC (J2000)}              & Declination of source. \\
4&\texttt{MAG\_AUTO\_CORR}.        & Automatic F814W magnitude from \textsc{SExtractor} catalogue, corrected according to the prescription discussed in Paper II.  \\
5&\texttt{F814W (GF)} & F814W apparent magnitude of
\galfit\ model. No k- or extinction correction applied. \\
6&\texttt{$\sigma_{F814W}$ (GF)} &Error on F814W, as calculated from the
simulations described in the text.\\
7&\texttt{\Reff\ (GF)} & Effective radius, in pixels
from \galfit\ model. \\
8&\texttt{$\sigma_{\Reff}$ (GF)}& Error on {\Reff}, as calculated from the
simulations described in the text.\\
9&\texttt{$n$ (GF)} & S\'ersic index from \galfit\ model. \\
10&\texttt{$\sigma_n$ (GF)} &Error on $n$, as calculated from the
simulations described in the text.\\
11&\texttt{$\mu_{e}$ (GF)}.& Surface brightness at \Reff\ calculated from the
\galfit\ model.\\
12&\texttt{\mueff\ (GF)}& Mean surface brightness within \Reff\
calculated from the \galfit\ model.\\
13&\texttt{Ellip. (GF)}            & Ellipticity of model ($1-b/a$). Errors as given by {\galfit}. \\
14&\texttt{$\sigma_{Ellip}$ (GF)}&Formal error on ellipticity from {\galfit}.\\
15&\texttt{$\theta$ (GF)}          & Position Angle of source. Errors as given by {\galfit}. \\
16&\texttt{$\sigma_{\theta}$ (GF)}&Formal error on $\theta$ from {\galfit}.\\
17&\texttt{RFF(GF)}              & Residual Flux Fraction diagnostic,
defined in the text, derived from the residuals from the \galfit\ model. \\
18&\texttt{EVI (GF)}              & Excess Variance Index, derived from
the residuals from the \galfit\ model. \\
19&\texttt{F814W (G2D)} & F814W apparent magnitude of
\gdd\ model. No k- or extinction correction applied. \\
20&\texttt{$\sigma_{F814W}$ (G2D)} &Error on F814W, as calculated from the
simulations described in the text.\\
21&\texttt{\Reff\ (G2D)} & Effective radius, in pixels
from \gdd\ model. \\
22&\texttt{$\sigma_{\Reff}$ (G2D)} &Error on {\Reff}, as calculated from the
simulations described in the text.\\
23&\texttt{$n$ (G2D)} & S\'ersic index from \gdd\ model. \\
24&\texttt{$\sigma_n$ (G2D)} &Error on $n$, as calculated from the
simulations described in the text.\\
25&\texttt{$\mu_{e}$ (G2D)}& Surface brightness at \Reff\ in the
\gdd\ model.\\
26&\texttt{\mueff\ (G2D)}& Mean surface brightness within \Reff\
calculated from the \gdd\ model.\\
27&\texttt{Ellip. (G2D)}            & Ellipticity of model ($1-b/a$). Errors as yielded by {\gdd}. \\
28&\texttt{$\sigma_{Ellip}$ (G2D)}&Formal error on ellipticity from {\gdd}.\\
29&\texttt{Pos Ang. (G2D)}          & Position Angle of source. Errors as given by {\gdd}. \\
30&\texttt{$\sigma_{\theta}$ (G2D)}&Formal error on $\theta$ from {\gdd}.\\
31&\texttt{RFF(G2D)}              & Residual Flux Fraction diagnostic,
defined in the text, derived from the residuals from the \gdd\ model. \\
32&\texttt{EVI (G2D)}              & Excess Variance Index, derived from the residuals from the \gdd\ model. \\
\hline
\end{tabular}

\caption{Output parameters included in the structural parameter
  catalogue. }
\label{tab:cat_explanation}
\end{table}

\subsection{Surface brightness, magnitude and size limits}
\label{cut:mag}

Although the two codes were used to fit all the sources 
detected by \sex\ in the F814W ACS images, not all of the resulting output parameters are 
meaningful and we apply limits in radius, magnitude and surface
brightness to the final table, explained below.

The Monte Carlo simulations presented in subsections
\S\ref{g2d:simulations} and \S\ref{galfit:simulations} indicate that the uncertainty and reliability 
of the output parameters depend critically on the $S/N$ ratio of the
fitted sources. This is in turn a function of magnitude and surface
brightness. Based upon these simulations we find that the derived
parameters and their errors are reliable for all sources for which:
$F814W < 24.5$,  and $\mueff\ < 24.5$.

Most galaxies for which $F814W \geq 24.5$ are likely
to be background galaxies, and in any case are beyond the magnitude limit of
current spectroscopic surveys, so cluster membership cannot be
verified, and their use in structural studies would be limited. 
At the distance of the Coma cluster, the apparent magnitude limit corresponds to
$M_{\mathrm{F814W}}=-10.5$, which is well within the absolute
magnitude distribution of globular clusters. 
 
We do find, however, that a number of galaxies with comparatively high
{\em central} surface brightness, large $R_\mathrm{e}$, and in some cases with measured
redshifts, fall below the surface brightness limit of $\mueff\ =
24.5$. For this reason, we also include in our catalogue galaxies with
$26.0 > \mueff\ > 24.5$, but we caution that because our simulations
largely did not cover the parameter space occupied by these galaxies,
we have less confidence in the derived structural parameters and their
errors.  For these sources for which only the \emph{central} regions are detected, both codes
are naturally forced to extrapolate a substantial part of the total surface brightness profile. In these cases, 
the results critically depend on the different hypothesis
(e.g. constant sky) with which the codes work,
and thus the derived parameters are more uncertain.

In the final catalogue we include only sources for which $F814W <
24.5$ as measured with {\em both} codes, provided that both codes had
converged. If only one code converged then the magnitude from that
code is used. For inclusion, sources have to satisfy the surface
brightness criterion for either of the codes, not both.

Besides magnitude and surface brightness cuts, we impose an \Reff\ lower limit
with the aim to eliminate point sources from the catalogue, since for
these the parameters of the S\'ersic fit have no meaning.
From the simulations described in Sect.~\ref{cut:re}, 
we have decided to reject sources for which either code
measures $\Reff < 1.1$ pixels, which is 2$\sigma$ above the mean of
the distribution in Figure ~\ref{fig:sim_logre_vs_mu}. A
number of point sources remain in the catalogue, in particular some bright, saturated
stars whose wings give a larger {\Reff}. Following visual inspection of
our images we estimate that $<$ 2\% of the objects in our sample are unresolved, these are a
mixture of unrejected stars, and unresolved objects such as Coma cluster
Ultra-Compact dwarfs. 

\begin{figure}
\includegraphics[width=8.5cm,angle=0]{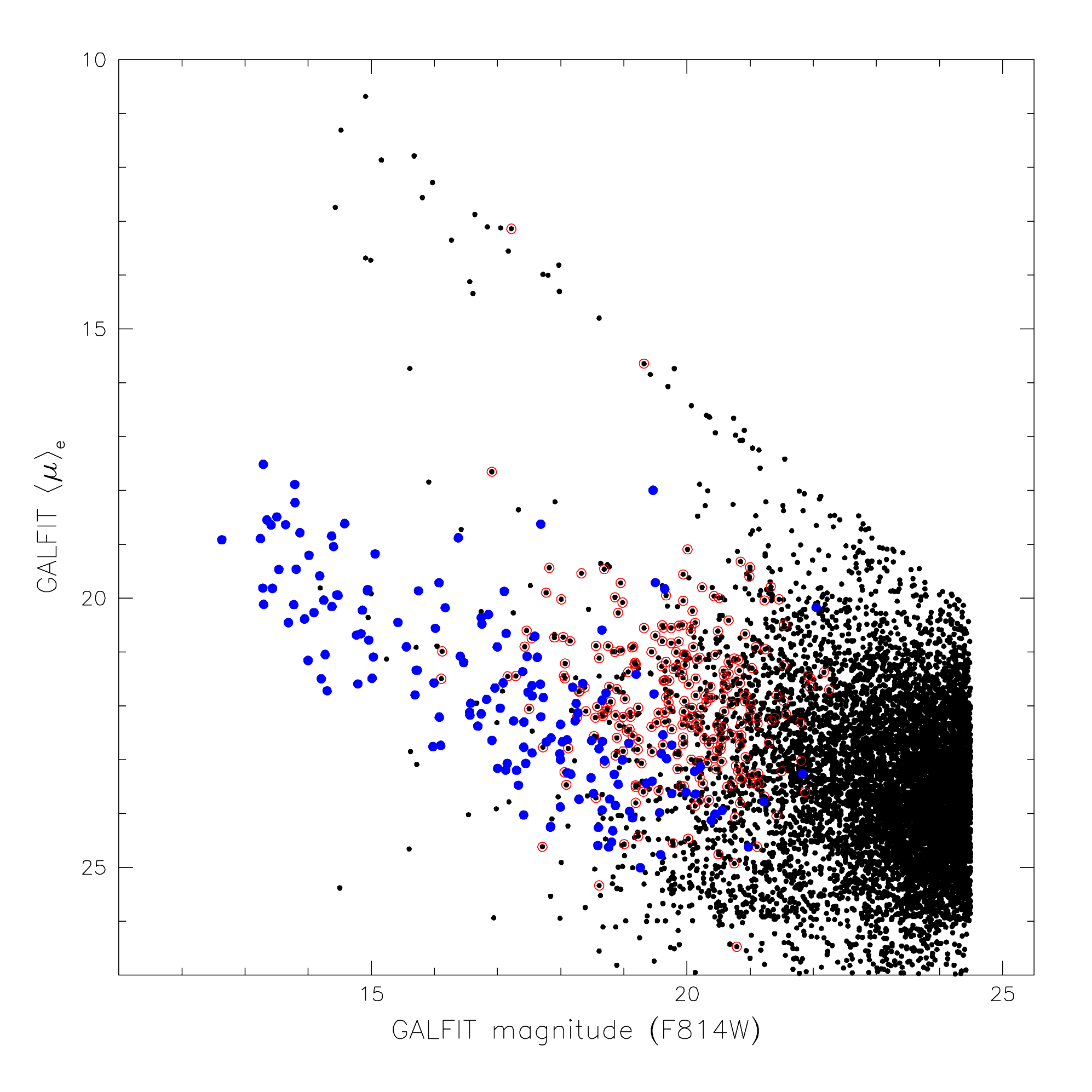}
\caption{\mueff\ plotted against F814W magnitude for the objects in the catalogue. Blue points are
spectroscopically confirmed cluster members, red circles are spectroscopically confirmed not to be members.
Black dots without a red circle are objects with no measured redshift. The diagonal line of black dots at upper right
is the remaining, partly saturated, stellar contamination. }
\label{fig:mag_vs_mu}
\end{figure}

\subsection{Magnitude-surface brightness relation}
\label{sec:mag-mu-plot}

Although we explicitly exclude any physical analysis of the structural catalogue in the present 
paper, we give a flavour of the types of objects included in the catalogue by 
showing (Fig.~\ref{fig:mag_vs_mu}) a plot of \mueff\ against magnitude, measured
with \galfit, for the objects in the catalogue. 
In this plot, blue denotes confirmed cluster members
(from the redshifts of Marzke et al. 2010), red circles confirmed non-members, and the black dots 
objects without measured redshifts. 

There is a diagonal line of black (and a few red) points at the upper right,
this is the residual stellar contamination, in the case of the bright points these stars are saturated in the 
ACS images, and hence have measured $\Reff > 1.1$ pixels. The cluster members form a sequence towards
the lower left, with a positive correlation between \mueff\ and luminosity. This sequence has considerable scatter
as the galaxies have not been selected by morphological type nor $n$. Confirmed non-members (almost all
background) have in general higher surface brightness for their apparent magnitude;
however, there are also cluster members with brighter {\mueff}.  These are compact ellipticals similar to those discussed by Price et al. (2009). 

\subsection{External Comparison}

\begin{figure*}
\begin{center}
\vspace{-2cm}
\includegraphics[scale=0.4]{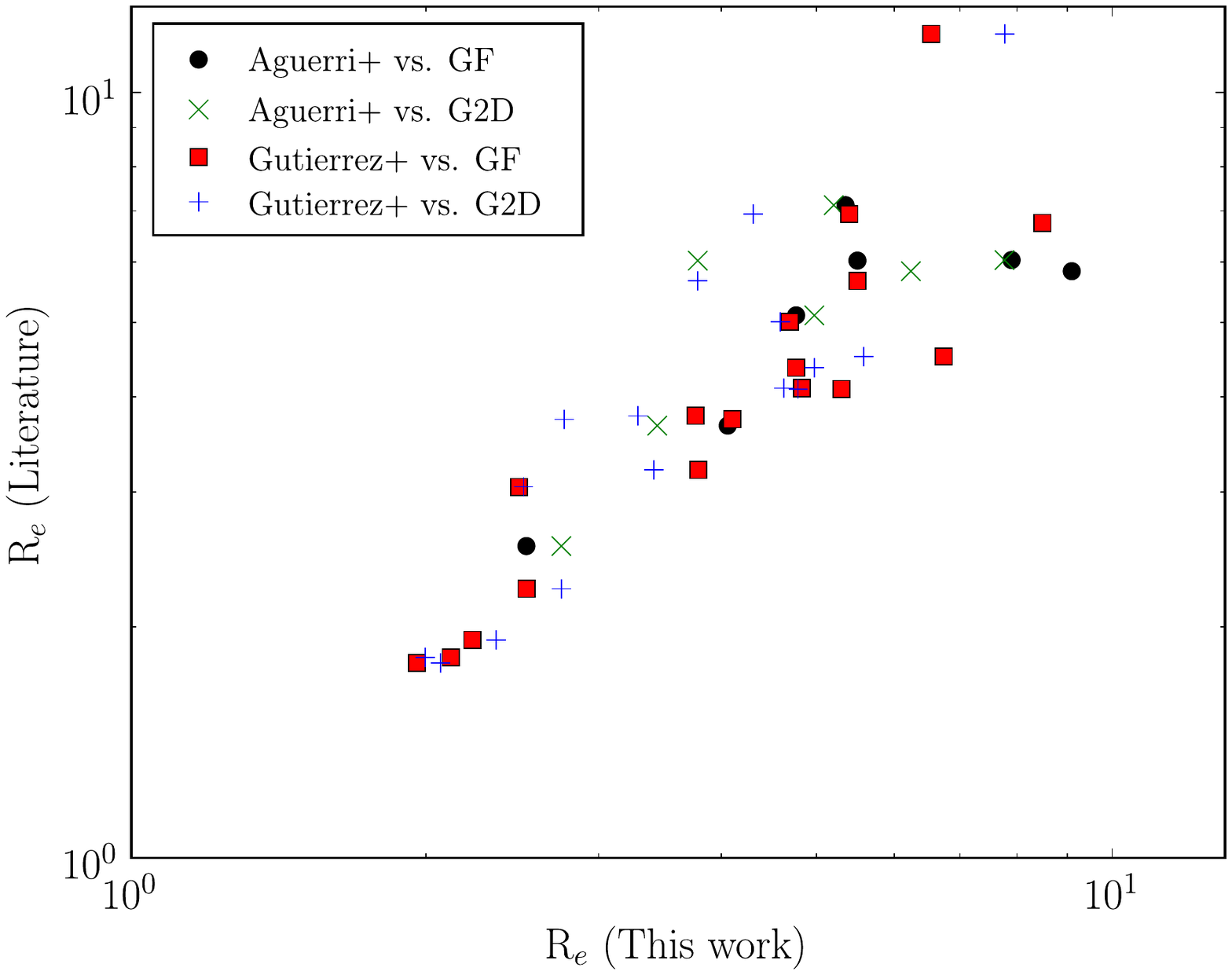}\includegraphics[scale=0.4]{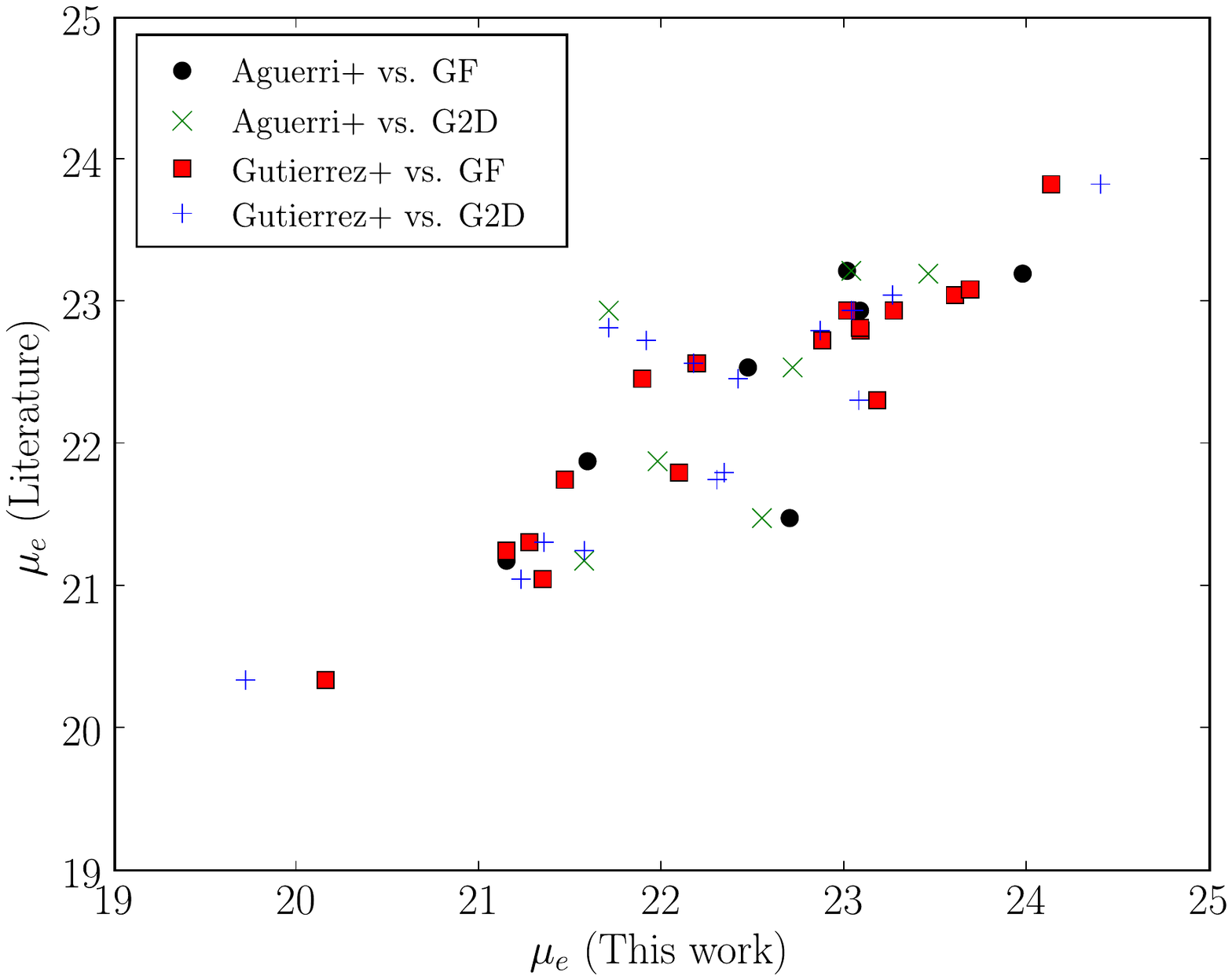}
\vspace{-2cm}
\caption{Comparison of our derived structural parameters with those of Aguerri et al.
(2005) and Gutierrez et al. (2004). Left panel: {\Reff} in arcseconds; Right panel $\mu_e$. Filled symbols
represent our \galfit\ values, and crosses our \gdd\ values, as described in the legend
at upper left in each panel.\label{fig:externalcomparison}}
\end{center}
\end{figure*}

{\bf Gutierrez et al. (2004) and Aguerri et al.(2005) study the structural parameters of dwarf galaxies 
in the Coma cluster using ground-based data obtained with the Isaac Newton Telescope at La Palma, 
in the Sloan-r filter. Some of the decompositions in these papers fit multiple components, here 
we only focus on the galaxies which have bulge-to-total ratios $B/T \geq 0.8$ in the Gutierrez et al. sample and 
the dEs (excluding the dS0s) from the Aguerri et al. sample. 17 galaxies from Gutierrez et al. and 7 galaxies
 from Aguerri et al. match within 2.0 arcseconds with galaxies in our structural catalogue. In 
Figure \ref{fig:externalcomparison} we present a comparison between the effective radii \Reff\ and $\mu_e$, the surface 
brightness {\bf at} \Reff, for these galaxies. The correlation is good, with a few outliers, mostly galaxies with complex 
structures in their centres which are not well resolved in the ground-based data. }

\section{Comparison between GIM2D and GALFIT}
\label{sec:comparison}

Both the \gdd\ and \galfit\ simulations have shown that these two codes are capable of recovering 
parameters of simulated galaxies in an almost unbiased way down to very low $S/N$ values. However, galaxies 
seldom consist of a single S\'ersic component. This makes a comparison between \galfit\ and \gdd\ for real sources 
interesting. Even though both codes try to minimise a Figure-of-Merit
(a $\chi^2$ value), there are significant
differences in both the way this minimisation is achieved (Markov Chain Monte Carlo \textsl{vs.} 
Levenberg-Marquardt minimisation) and how the best fit parameters are determined (median value 
of a number of realizations in the case of \gdd\ \textsl{vs.} final value after a number of iterations
in the case of \textsc{Galfit} ). These differences could, in principle, have an impact on the performance of
both codes even in the case of objects satisfying the magnitude, \mueff\ and \Reff\ cuts described
in \S~\ref{sec:catalogue}.

\subsection{Comparison of the fitted values}
\label{sec:compa1}

\begin{figure*}
\begin{center}
\includegraphics[scale=0.4]{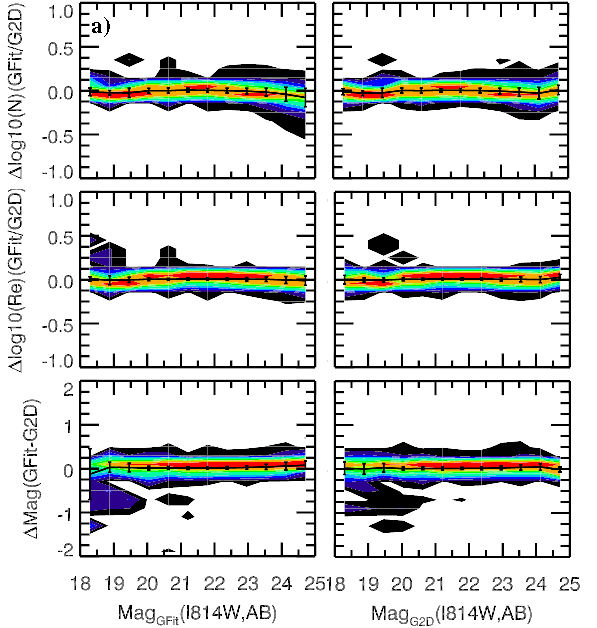}\includegraphics[scale=0.4]{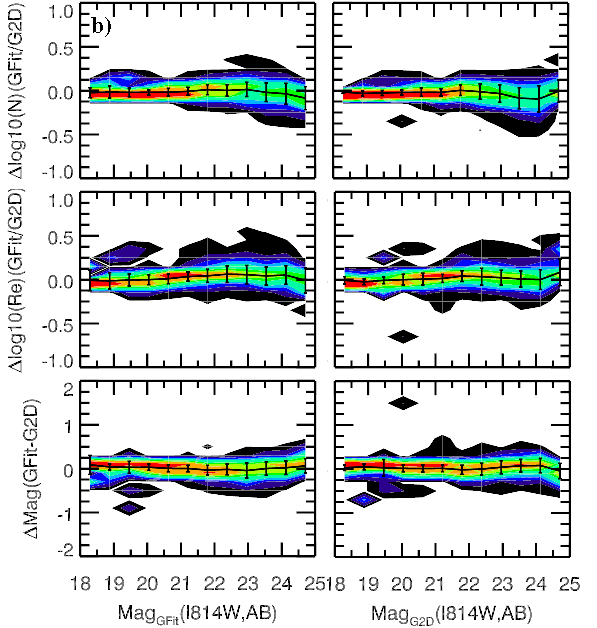}
\vspace{0.2cm}
\hspace{0.5mm}\includegraphics[scale=0.4]{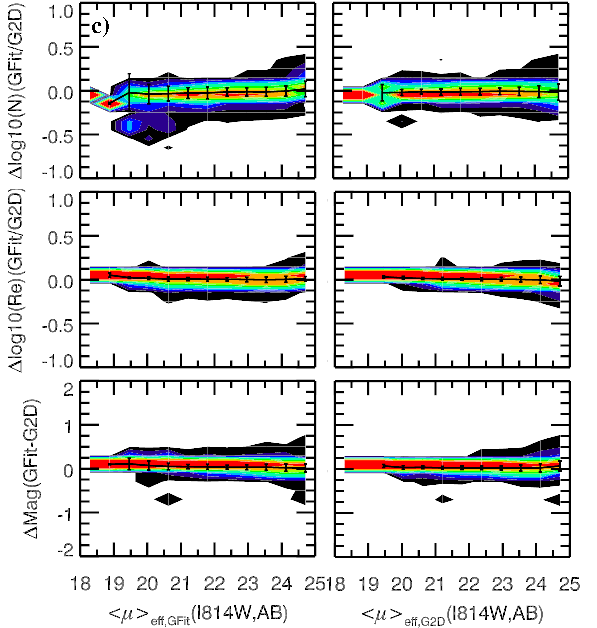}\includegraphics[scale=0.4]{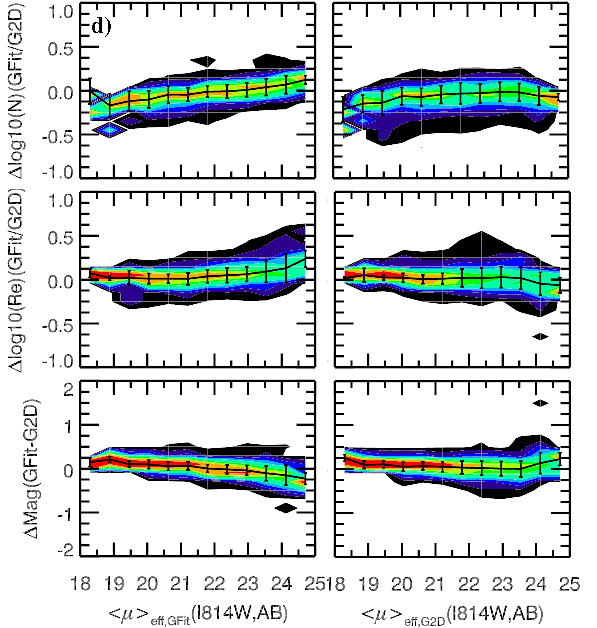}
\vspace{0.5cm}
\caption{Quadrant (a):  Magnitude, effective radius and S\'ersic
  index residuals as a function of 
\galfit\ magnitude (left panels) and \gdd\ magnitude (right
panels). This quadrant shows results for galaxies with $n < 2.5$. Quadrant (b): same for objects with $n > 2.5$.
Quadrant (c): Magnitude, effective radius and S\'ersic index residuals
as a function of \galfit\ \mueff\ (left panels) and 
\gdd\ \mueff\ (right panels). This quadrant shows results
for galaxies with $n < 2.5$. Quadrant (d): same for
objects with $n > 2.5$. Details of the vertical
  error bars and colour coding, are as in
  Fig.~\ref{fig:g2d_simulations}.\label{fig:gf_g2d_comp}}
\end{center}
\end{figure*}

The top two quadrants of Fig.~\ref{fig:gf_g2d_comp}  show the difference between
the measured magnitudes, effective radii and S\'ersic indices plotted against
the measured F814W magnitude. Quadrant (a) presents the results
for objects with a measured S\'ersic index lower than 2.5, whilst
quadrant (b) presents the results for objects of higher S\'ersic indices.
Quadrants (c) and (d) of this figure  present these differences plotted against the 
F814W \mueff\ for the same division in S\'ersic index.
The objects included in these plots are a subset of the full catalogue
described in \S~\ref{sec:catalogue}. The additional constraints for
inclusion in these plots are:
\smallskip 

\begin{enumerate}
\item $\mueff\ \leq 24.5$ 
\item $\textrm{RFF}\leq0.11$  and $\mathrm{EVI}\leq0.95$
\item $0.5<n<8.0$
\end{enumerate}
\smallskip

In each case the galaxies had to satisfy the constraint for both codes.
These additional cuts guarantee that the objects included in these
plots are well measured and  well represented by S\'ersic
profiles. 

These plots show that, whenever the two codes yield a  S\'ersic index lower than 2.5, the observed 
agreement between the output parameters is quite remarkable even for the S\'ersic index 
itself. In this case, the expected magnitude scatter between the two codes is around
0.25 mag at the faintest levels, with a similar agreement in the \Reff\ measurements.

Quadrants (b) and (d) of Fig.~\ref{fig:gf_g2d_comp}  show that
in the cases in which the two codes measure a S\'ersic index higher than 2.5, the disagreement
between the two codes is larger. 
It is therefore more difficult to get consistent measurements
for objects with extended wings containing a large fraction of the object's flux. This 
disagreement between the two codes manifests itself not only as a larger scatter, but also as a 
systematic trend of the effective radii and S\'ersic index residuals, when plotted against the
measured surface brightnesses.

Whenever one code finds a brighter magnitude than the other, it also
finds \Reff\ to be larger. This is in agreement with Fig.~\ref{fig:g2d_simulations_1}. This is probably caused by differences in the
sky values adopted by the two codes. This is also
the probable cause of the poorer agreement between the two codes when fitting objects with high S\'ersic indices.
For these objects, and in particular for sources that are found to be in the luminous haloes of the largest galaxies, it
is very difficult to define the sky level with accuracy simply because these sources have a large
fraction of their total flux in very extended wings and thus, any small discrepancy in the measured sky level
translates into a large difference in the final luminosity and effective radius.

Fig.~\ref{fig:gf_g2d_comp} also shows that the scatter in the output S\'ersic index
is larger for higher values of this index. This is caused by the fact that models
of high S\'ersic index start to converge towards a limiting
$r^{-2}$ profile and it is therefore very difficult to distinguish
between two such models.

Another discrepancy between the two codes comes from small sources. Often these sources are fit 
by \galfit\ with a very high Sers\'ic index, and a half-light radius of a few pixels. These fits have a substantial
portion of their light in wings which are below the detection threshold. On the other hand, \gdd\ fits these 
sources with lower S\'ersic indices, and the \gdd\ fit is probably a more realistic representation of the detected
image. 

One advantage of \galfit\ is that it can model a number of galaxies simultaneously, leading to more consistent 
fits in cases where the target galaxy has near neighbours. \gdd\ sometimes has  problems with these sources as
the only thing \gdd\ can do with neighbouring sources is to mask them, and that fraction
of the flux from these neighbouring sources which falls outside of their masks will compromise the fits.  

\section{Summary, conclusions and future work}
\label{sec:conclusions}

We have fitted single S\'ersic models to the surface brightness distributions of 74992 galaxies included in the 
photometric catalogue presented in Paper II, using the two most widely used two dimensional
galaxy fitting codes, \gdd\ and {\galfit}. Both codes create a PSF convolved trial
function which is then compared to the input science data. The PSFs
used were created mimicking the most important reduction steps that 
the science images had undergone.

Independent simulations show that both codes can achieve 
similar accuracies for most cases. However, \gdd\ requires a 
much higher degree of intervention compared with \galfit\ to produce the
results. This is due to the greater 
flexibility of {\galfit}, which makes it easier for any given input image to
satisfy the fitting hypothesis under which \galfit\ works. \gdd\
needs more manual work, including the use of enlarged masks with sizes 
calculated according to our detailed
Sextractor simulations, and the masking of nearby sources, to produce reliable results.

These simulations, together with an additional constraint aimed at rejecting probable point sources have been used 
to select a subsample of galaxies from the initial Sextractor
catalogue. The criteria used are:

\begin{itemize}

\item $\mueff\ < 26.0$, where \mueff\ is the mean surface brightness
enclosed within {\Reff}.
\item $F814W < 24.5$
\item \Reff\ $>$ 1.1 pixels
\end{itemize}

We find that 8814 objects satisfy these criteria, cross-matching with
the redshift catalogue of Marzke et al (2010), indicates that 424 of
these have redshifts, of which 163 have redshifts which place them as probable cluster members.

We have also introduced two different goodness of fit diagnostics. The
Residual Flux Fraction is an indicator of the
amount of light contained in the residuals which cannot be 
accounted for by the per-pixel photometric errors associated with the science
image. The Excess Variance Index measures the importance of the internal
structure present in luminous galaxies that the S\'ersic model is 
unable to reproduce. These diagnostics indicate 
whether the surface brightness distributions of the fitted galaxies 
are well described by the S\'ersic profiles, or alternatively  whether
it is valuable to use more complex functional forms (maybe adding a nuclear point source, or
a bulge+disk decomposition) than the simple S\'ersic profile used in
this work. 
 
Although the S\'ersic fits presented here provide a good overall
description of the majority of galaxies in our sample, the spatial
resolution and stability of the PSF provided by HST and ACS allow the
surface brightness to be measured accurately over a wide dynamic
range. A range of more complex functions have been fit to surface
brightness profiles, such as the ``dual power law''
(Ferrarese et al.~1994), the ``Nuker Law'' (Lauer et al.~1995), and
the core-S\'ersic profile (Graham et al.~2003; Trujillo et
al. 2004).  In the context of elliptical and spheroidal galaxies, a number of
authors (e.g. Graham and Guzm\'an 2003; Graham 2004; 
Ferrarese et al.~2006; C\^ot\'e et al.~2007; Merritt \&
Milosavljevi{\'c} 2005; Kormendy et al.~2009), have identified and
measured either ``extra light'' or ``missing light'' with respect to
the S\'ersic formalism in their profiles. 
In future papers in this series we will investigate in detail these
deviations from the S\'ersic profile, and the correlation with kinematic properties
(Emsellem et al.~2007; Cappellari et al.~2007; Krajnovi\'c et al.~2008, van Zee et al. 2004,
Toloba et al. 2009).  

An alternative formalism which we will test for surface brightness profiles of
ellipticals derives from the modified isothermal models of King (1962,
1966) and Wilson (1975), which are very successful 
in reproducing the surface brightness distributions of star clusters,
and of some, but not all, elliptical galaxies. 

In the current paper we analyse only the radial surface brightness
profile, but there is also information in the azimuthal distribution,
and we will undertake decompositions into disk, bulge, and nuclear
components where appropriate (Andredakis et al.~1995; Balcells et al.~2007a,b), which will
permit the construction of more meaningful structural
parameter diagrams for the different physical components.

We will also investigate asymmetric deviations from the
basic S\'ersic formulation of the light distribution, using techniques for
parameterising galaxy surface brightness such as CAS (Conselice 2003) and
Gini/M20 (Lotz et al.~2004), and for detecting and quantifying bars
(Marinova et al.~2010).

\section*{Acknowledgments}

We thank Dr. Luc Simard for providing the code \textsc{DrizzlyTim}. DC and AMK acknowledge support from the Science and Technology
Facilities Council, under grant PP/E/001149/1. CH acknowledges financial support
from the Estallidos de Formaci\'on Estelar. Fase III, under grant AYA
2007-67965-C03-03/MEC. MB acknowledges financial support from grants AYA2006-12955 and AYA2009-11137 from the Spanish 
Ministerio de Ciencia e Innovaci\'on. DM was supported by grants AST-0807910 (NSF) and NNX07AH15G (NASA).

\appendix

\section{Comparison to GIM2D and GALFIT results by the GEMS collaboration.}
\label{apx:g2d_gfit07}

Wanting to build upon the experience obtained by other teams in using GALFIT and GIM2D on HST/ACS data, we decided that the best starting point is the extensive GALFIT/GIM2D comparison by the GEMS collaboration (H\"aussler et al. 2007) on mock HST/ACS observations. The GEMS collaboration provides public access to the analysis and actual data on
the web (\texttt{http://www.mpia.de/GEMS/fitting\_paper.html}). We used their \texttt{Bulge0001} and \texttt{Disk0001} simulated observations. These contain artificial galaxies with  pure S\'ersic profiles with $n=4$ or $n=1$
respectively, with appropriate Poisson noise, on top of a noise background created by putting together patches of real sky observations from different images.
These simulated images are created for the F850LP filter using a PSF with a FWHM typical of this filter. The artificial galaxies cover a significant fraction of the galaxy parameter space expected for the Coma HST/ACS data, in terms of $S/N$ ratio, total magnitude and effective radius. Therefore our analysis of these data is useful even though the Coma Legacy Survey does not use the F850LP filter.

Following the work by H\"aussler et al. (2007) we fit a pure S\'ersic model with a flat background. All the parameters, including the
S\'ersic index, were allowed to vary freely, although the value of the residual sky was later fixed, as explained below.
The point spread function used was identical to the one used to create the mock galaxies.

For \galfit\ we converged on a very similar fitting setup and best fitting results as H\"aussler et al..
For \gdd\ we arrived at a different setup and masking treatment which led to improvements in the best fitting results.

First, \gdd\ was run with its default decision algorithm on the size of the cutout to be fitted. It was also allowed to fit
and refine the sky value by itself (\texttt{DOBKG=YES}) and to automatically estimate initial values for the fitted parameters (\texttt{INITPARAM=YES}).
The S\'ersic index was left as a free floating parameter. The \sex\ parameter configuration ensures detection of sources with a global signal-to-noise ratio of 20.25. The \sex\ parameter \texttt{BACK SIZE} was set to 512. The minimum stamp size was set to 31px. This experimental configuration leads to the ``recommended'' setup for {\gdd}.
The noise model used for the \gdd\ fits was based on the traditional CCD equation using the background $\sigma$ and the image effective gain which
is ultimately regulated by the exposure time and reduction process.

\begin{figure}
\includegraphics[scale=0.45]{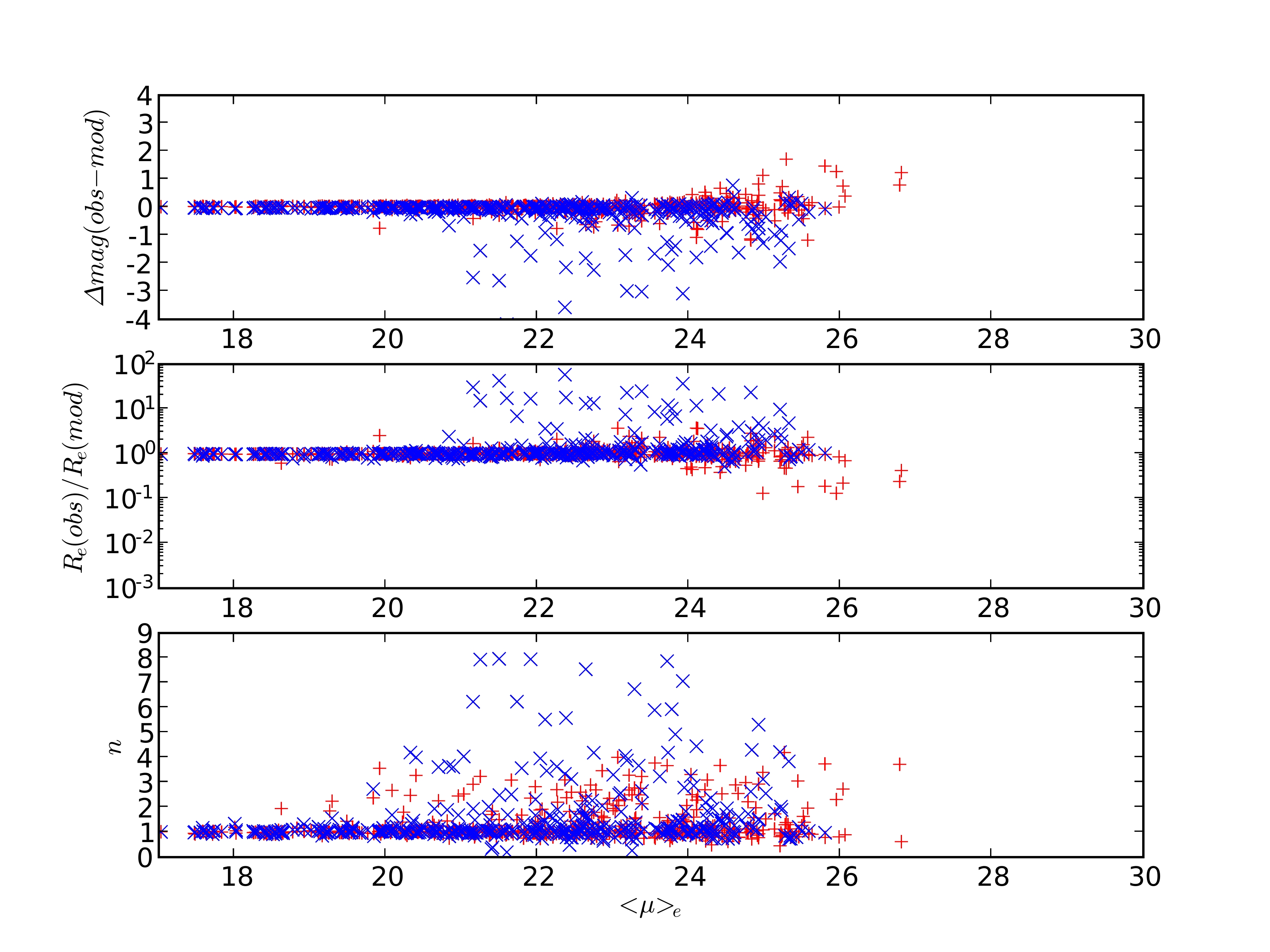}
\caption{Results for GIM2D on pure exponential models. From top to bottom: magnitude residual (GIM2D minus Model), ratio of effective radius (\gdd\ / Model) and S\'ersic index as a function of the true average surface brightness within an effective radius. Red crosses are our GIMD2D runs, blue crosses are the results for the same simulated galaxies from the GEMS team.}
\label{gems_g2d_disks_1}
\end{figure}

For the Disk0001 models \gdd\ produced a fit for 470 simulated profiles. Figure \ref{gems_g2d_disks_1} shows the
results obtained by \gdd\ in this case. The upper panel presents the magnitude residuals, the middle panel presents the ratio in effective
radii, and the lower panel presents the output S\'ersic index as a function of the average surface brightness within a effective radius of
the simulation. Our \gdd\ results are shown by red crosses and the GEMS team results with \gdd\ by the blue circles. The general conclusion from this is that there are
systematic trends at the faintest surface brightness levels, fainter than $\mueff \geq 22.5$. The scatter between fit and simulation
is smaller for our \gdd\ run compared with the results of the GEMS team. Likely,this is caused by the fact that we masked overlapping sources with
a value of -2 in the mask image, a situation that forces \gdd\ to ignore those problematic pixels. The GEMS group did not do this.

\begin{figure}
\includegraphics[scale=0.45]{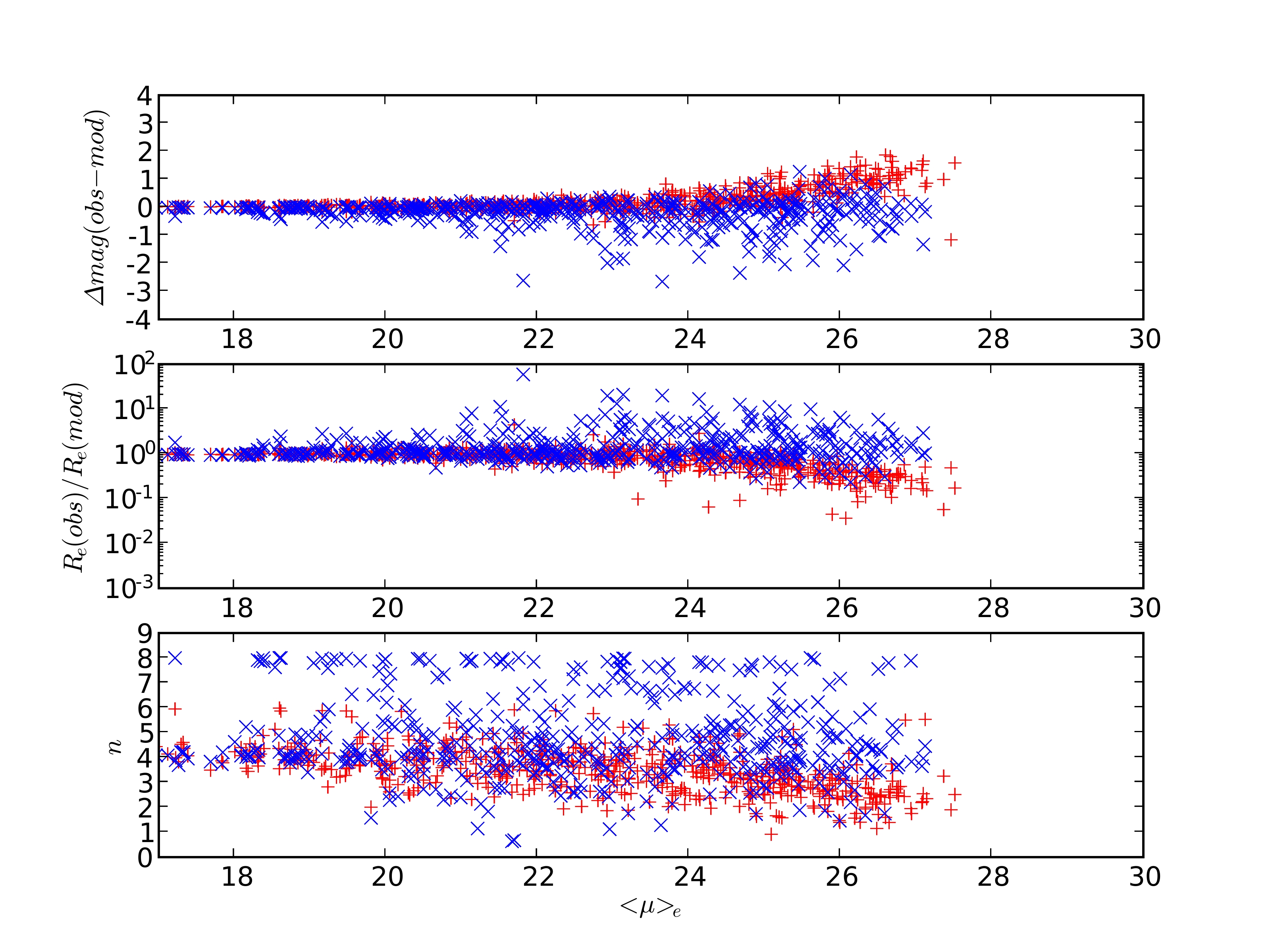}
\caption{Results for GIM2D on pure de Vaucouleurs models. From top to bottom: magnitude residual (GIM2D minus Model), ratio of effective radius (\gdd\ / Model) and S\'ersic index as a function of the true average surface brightness within an effective radius. Red crosses are our GIMD2D runs, blue crosses are the results for the same simulated galaxies from the GEMS team.}
\label{gems_g2d_bulges_1}
\end{figure}

For the Bulge0001 models \gdd\  yielded a successful fit for 558 simulations. Figure \ref{gems_g2d_bulges_1} shows the results. The general conclusion is that our results are an improvement on the GEMS results down to a surface brightness of around $z\sim22.5$, and
have a smaller scatter. However there are again systematic trends which increase with fainter surface brightness.

We investigated if the systematic deviations shown by the \gdd\ solutions at surface brightness levels fainter
than $z=23.0$ could be due to an inaccurate sky. To this end a number of low surface brightness sources were refit. This time, we used a sky determination derived
from a customized configuration file for \sex\, whose main difference with the \sex\ configuration file used in the previous
fitting round is that it uses a 2048 mesh for background subtraction and determination. Also, we did not allow \gdd\ to fit the sky. Furthermore the cutout size for \gdd\ was increased to 25 times the default size.

\begin{figure}
\includegraphics[scale=0.45]{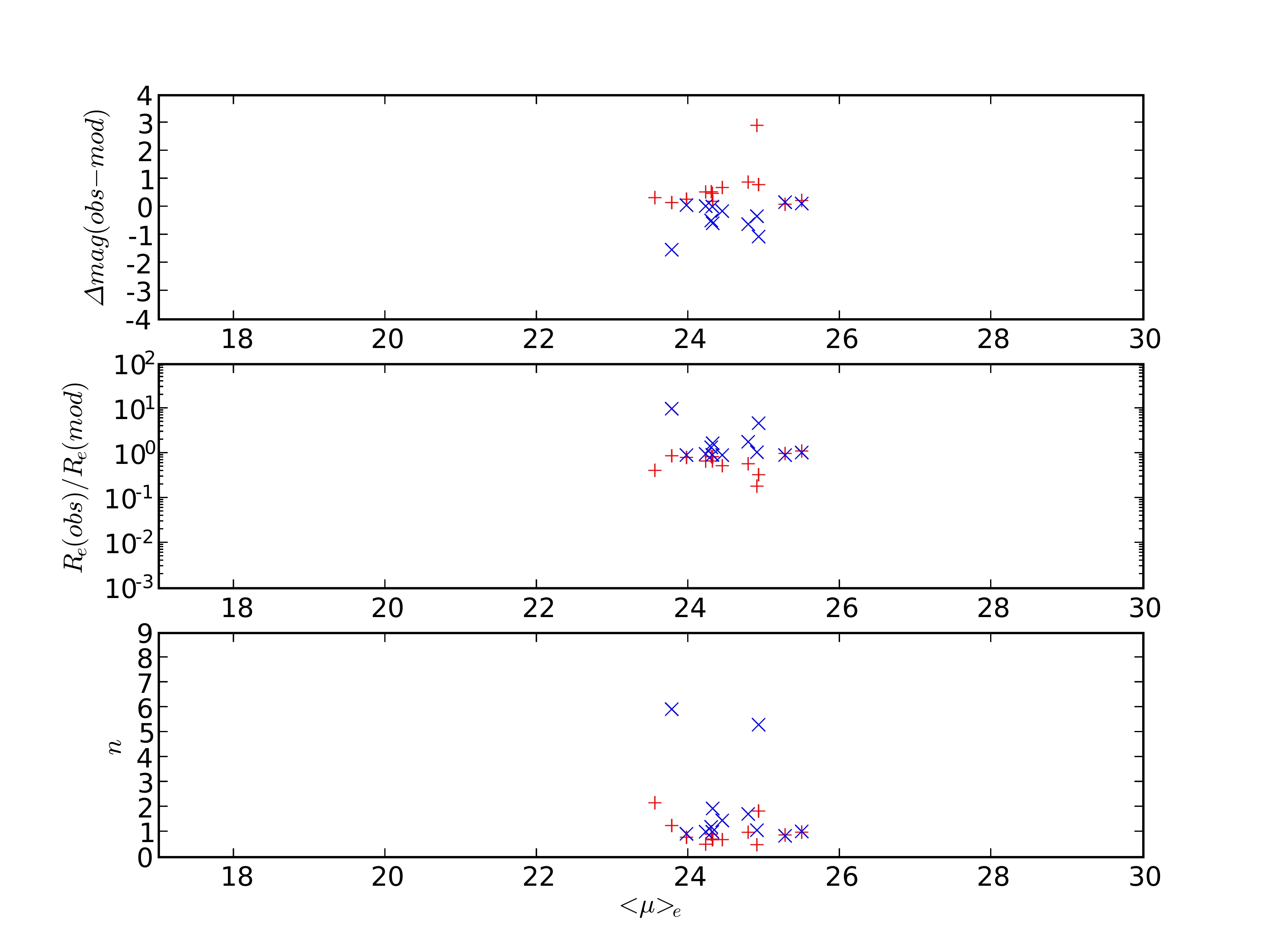}
\caption{Results for GIM2D on pure exponential models of low surface brightness, using our own sky determination (see text). From top to bottom: magnitude residual (GIM2D minus Model), ratio of effective radius (\gdd\ / Model) and S\'ersic index as a function of the true average surface brightness within an effective radius. Red crosses are our GIMD2D runs, blue crosses are the results for the same simulated galaxies from the GEMS team.}
\label{gems_g2d_disks_3}
\end{figure}

Figure \ref{gems_g2d_disks_3} shows the
result of this test for the mock galaxies present in image \texttt{Disk0001}. Comparison with Figure \ref{gems_g2d_disks_1} shows that, using
this prescription to deal with the sky level, \gdd\ no longer shows systematic offsets in magnitudes and effective radius even at the
faintest surface brightness levels.

\begin{figure}
\includegraphics[scale=0.45]{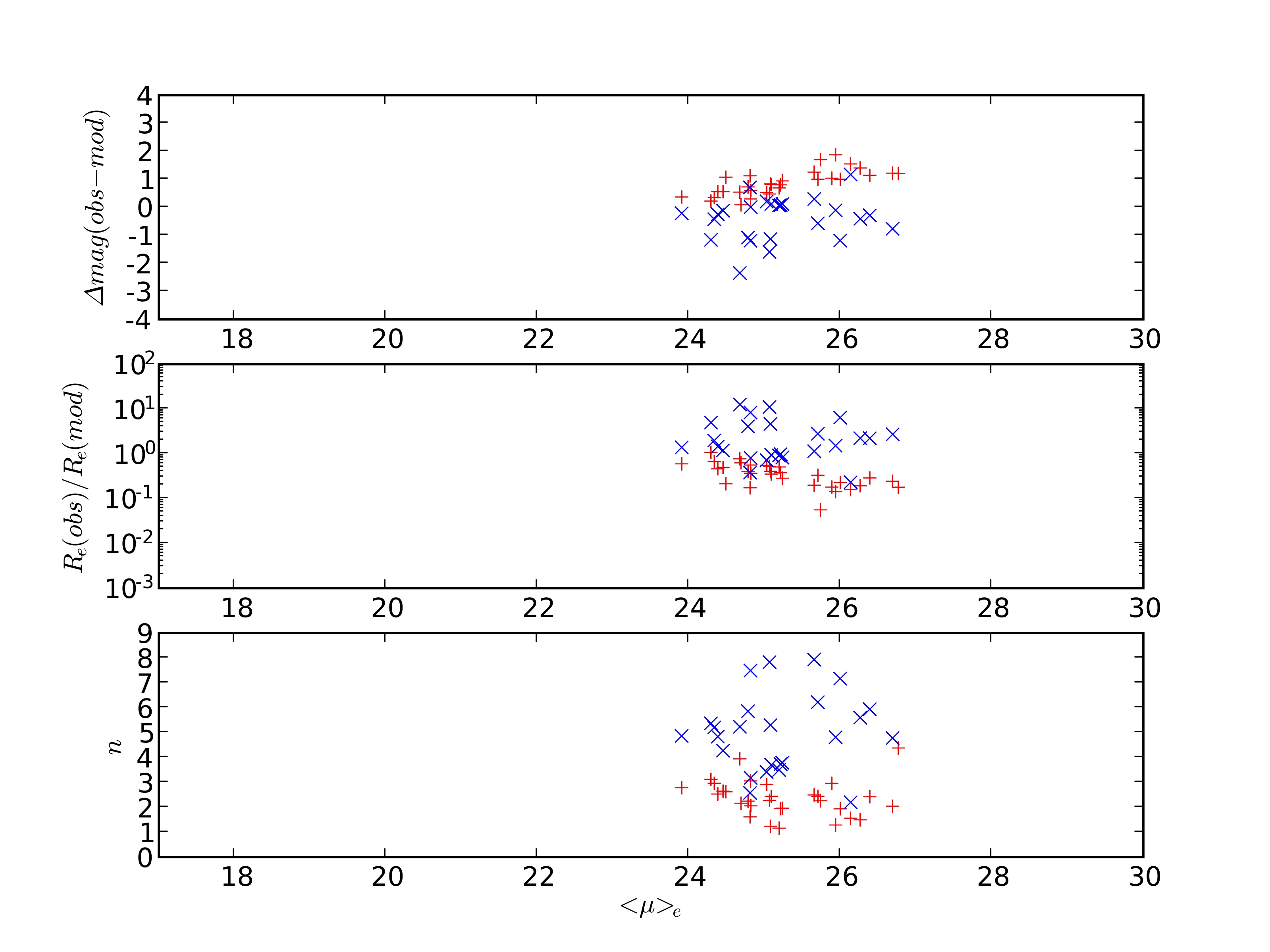}
\caption{Results for GIM2D on pure de Vaucouleurs models of low surface brightness, using our own sky determination (see text). From top to bottom: magnitude residual (GIM2D minus Model), ratio of effective radius (\gdd\ / Model) and S\'ersic index as a function of the true average surface brightness within an effective radius. Red crosses are our GIMD2D runs, blue crosses are the results for the same simulated galaxies from the GEMS team.}
\label{gems_g2d_bulges_3}
\end{figure}

Figure \ref{gems_g2d_bulges_3} shows the result of this fitting round for the models included in the \texttt{Bulge0001} image. Comparison
to Figure \ref{gems_g2d_bulges_1} clearly shows that setting the sky to be equal to the sky value given by \sex\ makes the systematic offsets
in magnitudes and effective radius to vanish even at the faintest surface brightness levels. The statistical errors (the scatter) are rather large however.

From the fits to the synthetic models generated by the GEMS team comprising either pure exponentials or pure de Vaucouleurs models, we conclude that:

\begin{itemize}
\item{The errors in the recovery of parameters appear to be dominated by errors in the sky determination.}
\item{Our G2D fits show significantly reduced scatter with respect to the G2D fits published by GEMS team. The difference can probably be traced to the manual masking of intervening sources in the object's mask done by our team.}
\item{Increasing the cutout size used by \gdd\ reduces or removes the systematic offsets in magnitude and effective radius found when the
default cutout size is used.}
\item{Overall, we conclude that we can use both the GALFIT and GIM2D codes with a similar or better accuracy to that found by the GEMS collaboration.}
\end{itemize}

\section{Creating object masks for GIM2D}
\label{apx:mask_recipe}

We provide a detailed recipe for producing object masks for \gdd\ that solves the missing flux problem that appears when standard \sex\ segmentation masks are used in \gdd. 

\begin{enumerate}

\item The aperture for the target  should ideally be equal to the Kron aperture of an object 
with the same \texttt{FLUX\_RADIUS} and $n$ that would be measured if the
observations had infinite $S/N$ ratio. The elongation and position angle
of this aperture would ideally be equal to those of the real object. The pixels within this aperture 
should be flagged as 1 in the mask file so that \gdd\ recognizes them as belonging to its target object.
This aperture is calculated from the \sex\ measurements.
\texttt{FLUX\_RADIUS}, \texttt{THETA\_IMAGE} and \texttt{ELONGATION} were taken directly from the \sex\ catalogue.
\texttt{MUOBS} is a \sex\ based estimate of effective surface brightness of that particular detection, obtained
as if \texttt{MAG\_AUTO} were the total magnitude and the 50\% of \texttt{FLUX\_RADIUS} were the real effective 
radius. It is thus an estimate of the effective surface
brightness. The 5th degree polynomials used in the \sex\ simulations 
presented in Paper II were then used to estimate empirically
the effective radius of the source. This effective radius estimate is then 
converted into a Kron aperture using the relationship between \Reff\ 
and $R_{\mathrm{1}}$ (the $S/N=\infty$ Kron aperture in units of the effective radius) and $n$.
$n = 2.5$ is assumed where $n$ is undefined.
Although the aperture defined in this way leaves out a finite fraction of the total flux from the 
target models outside the mask, this magnitude offset is 0.1 mag in the worst case. For this reason, the initial flux 
estimate will be around 0.1 mag fainter than the true integrated flux. As above, a very similar statement 
can be made about the initial effective radii estimates. Using this customized mask, \gdd\ has a 
realistic opportunity to measure 100\% of the target object total flux
and size, whereas with the default \sex\ segmentation image, this is not true.

\item The apertures for the background sources are elliptical apertures, with an area given by the
\texttt{ISOAREA\_IMAGE} \sex\ measurement. The elongation and position angles of these ellipses are also taken from 
the \sex\ measurements. These pixels are flagged as -2 so that \gdd\ ignores these pixels in all 
calculations. \gdd\ is therefore blind to the innermost 
regions of the background sources. Thus only the outer regions of the background sources contribute to the sky 
level affecting the galaxy under study. In the case of an overlap between target and
background object pixels, the pixels are assigned to the background
object as including them in the target would potentially compromise
the fit.

\item Finally, pixels not belonging to the target  or to the
 background sources are flagged as 0, so that \gdd\ regards these pixels
as sky.

\end{enumerate}

In a few cases, for instance, with objects with close neighbours or
near to the CCD edges, \sex\ did not provide measurements of \texttt{MAG\_AUTO} and 
\texttt{FLUX\_RADIUS}. In these cases \texttt{MUOBS} is estimated
by spreading the light encircled in the isophotal area of the 
object (\texttt{ISOAREA\_IMAGE}) uniformly  in an aperture with a
radius equal to $\frac{1}{2}\times\sqrt{\frac{\mathrm{ISOAREA\_IMAGE}}{\pi}}$. 
This was found to be somewhat brighter than a measured~ \texttt{MUOBS}, although
the final apertures obtained were similar in size.

The size of the poststage stamp over which \gdd\ has to work is 
an integral part of the suggested solution. 
The quantity \texttt{FRAD\_MOD} amounts, in the vast majority of the cases,
to 50\% of \texttt{FLUX\_RADIUS}, as calculated by {\sex}.
For the objects where \sex\ could not calculate \texttt{FLUX\_RADIUS},
\texttt{FRAD\_MOD} is half of the radius of the circle 
with an area equal to \texttt{ISOAREA\_IMAGE}, which always does exist.

The number \texttt{YFX(MUOBS)} is given by the empirical relation
between the 50\% \texttt{FLUX\_RADIUS} and the \emph{input} effective radius that
is appropriate for sources with a high $n$.  For reasonable S\'ersic indices
\texttt{YFX(MUOBS)} is given by the fifth-order polynomial in the
mean observed surface brightness \texttt{MUOBS}, presented in paper II. 
The final size of the post stamps imagelets 
extracted is $7.0\times\mathrm{FRAD\_MOD}\times\mathrm{YFX(MUOBS)}$ 
on a side.

\label{lastpage}

\end{document}